\RequirePackage{snapshot}
\documentclass[a4paper, 12pt]{article}

\usepackage[english]{babel}
\usepackage[utf8x]{inputenc}

\usepackage{booktabs}
\usepackage{tabu}
\usepackage[T1]{fontenc}
\usepackage{enumitem}

\usepackage[a4paper, margin = 1in]{geometry}

\usepackage{amsmath}

\usepackage{graphicx}
\usepackage{setspace}
\usepackage{verbatim}
\usepackage{amsfonts}
\usepackage[colorinlistoftodos]{todonotes}
\usepackage[colorlinks=true, allcolors=blue]{hyperref}
\usepackage[capposition=top]{floatrow}
\usepackage{caption}
\usepackage{subcaption}
\usepackage[flushleft]{threeparttable}
\usepackage{natbib}
\usepackage{appendix}

\newtheorem{assumption}{Assumption}
\newtheorem{theorem}{Theorem}
\newtheorem{lemma}{Lemma}

\title{Combining Observational and Experimental Data to Improve Efficiency Using Imperfect Instruments}
\author{George Z.  Gui \thanks{Columbia Business School, \href{mailto:zg2467@gsb.columbia.edu}{zg2467@gsb.columbia.edu}.
I am extremely grateful to my advisor Harikesh Nair  and Navdeep Sahni for their guidance and support. I am also greatly indebted to Robyn Lockwood for her writing instruction. I also thank a group of researchers whose comments have greatly improved the paper: Jacob Conway, Jacob Dorn, Wesley Hartmann, Guido Imbens,  Yewon Kim, James Lattin, Sridhar Narayanan, Peter Reiss, Evan Rosenman, Jann Speiss, and Raluca Ursu, and seminar participants at the Stanford Marketing WIP and the Econometrics Workshop. All errors are my own.}}
\date{September,  2023\\
\;\\
This is the final working paper version. Final published version may be found \href{https://pubsonline.informs.org/doi/abs/10.1287/mksc.2020.0435}{here},  Marketing Science}
\begin{document}

\maketitle

\begin{abstract}
Randomized controlled trials generate experimental variation that can credibly identify causal effects, but often suffer from limited scale, while observational datasets are large, but often violate desired identification assumptions. To improve estimation efficiency, I propose a method that leverages imperfect instruments - pretreatment covariates that satisfy the relevance condition but may violate the exclusion restriction.  I show that these imperfect instruments can be used to derive moment restrictions that, in combination with the experimental data, improve estimation efficiency. I outline estimators for implementing this strategy, and show that my methods can reduce variance by up to 50\%; therefore, only half of the experimental sample is required to attain the same statistical precision. I apply my method to a search listing dataset from Expedia that studies the causal effect of search rankings on clicks, and show that the method can substantially improve the precision.
\end{abstract}

\section{Introduction}
In digital marketing systems, firms often use algorithms to determine marketing variables such as prices, ads, search results, and product recommendations. Accurately estimating how these marketing variables causally affect subsequent outcomes, such as clicks and purchases, is particularly important for firms; otherwise, it is challenging for firms to  improve and optimize their algorithms.

However, these causal effects cannot be easily estimated using observational data because the estimates are often biased and inconsistent  (\citealp{lewis_here_2011}; \citealp{blake_consumer_2015}; \citealp{ursu_power_2018}; \citealp{gordon_comparison_2019}). These biases and inconsistencies are caused by endogeneity - that there exist factors that influence both the marketing variables and outcomes. To obtain an unbiased estimate, firms often proceed by running an experiment and making their future decisions based on the experiment results. Because experimentation could hurt user experience and profit, experiments often have limited scale, being conducted only within a subset of traffic, markets, or weeks. For example, Google \citep{wang_learning_2016}, Microsoft \citep{li_counterfactual_2015}, and Expedia \citep{ursu_power_2018} have all conducted experiments that randomized search results rankings for a small subset of traffic. Uber \citep{chen_value_2019} has conducted experiments that randomized incentives within one county for three weeks. Advertisers sometimes randomize ad exposure for several weeks to measure advertising ROI \citep{lewis_unfavorable_2015}. These experiments sometimes produce imprecise estimates due to their limited scale. To illustrate, \cite{lewis_unfavorable_2015} examine $25$ advertising field experiments and document that the confidence interval on advertising ROI is wider than $100\%$, implying that advertisers cannot distinguish a campaign with a low $0\%$ ROI from a campaign with a highly profitable $50\%$ ROI. Therefore, it is particularly important to develop methods that improve the estimation efficiency.

This paper proposes a method that combines experimental and observational data to improve estimation efficiency. The method hinges on having a set of imperfect instruments --- pretreatment covariates that satisfy the first-stage relevance condition in the observational data but may violate the exclusion restrictions \citep{nevo_identification_2010}. A common challenge of leveraging these imperfect instruments is that, due to the violation of exclusion restrictions, the corresponding observational estimate is biased and inconsistent. I demonstrate that this bias can be corrected using experimental data. More importantly, I show that this bias-corrected estimate is uncorrelated with the experiment-only benchmark estimate. Using weighting or GMM, these two uncorrelated estimates can be combined into a more precise estimate without introducing significant bias. I illustrate that the efficiency gain depends on the size of the observational data and how well the pretreatment covariates predict the treatment assignment.  When the observational data are large and the pretreatment covariates can accurately predict treatment assignment, I prove that my method can reduce the variance by up to 50\%.

The method is particularly useful in the digital marketing system because firms usually have access to a large observational dataset generated by an algorithm. The algorithm often takes many covariates as inputs to endogenously determine the treatment of interest. For example,  advertisers use customer characteristics to make targeting decisions,  Expedia uses hotel ratings and other hotel characteristics to rank different hotels,  and Uber uses supply and demand forecasts to determine how to charge customers and compensate gig workers. Due to the algorithmic nature of the data, any relevant inputs to the algorithm can be viewed as imperfect instruments. Therefore, if researchers observe inputs to the algorithm, or observe any other pretreatment covariates that may correlate with the inputs, researchers can use these inputs or covariates as imperfect instruments to improve the estimation efficiency.

Formally, the method requires two datasets, the experimental sample and the observational sample. For each unit in both samples, we observe the outcome of interest, the treatment, and a set of pretreatment covariates. The only difference between the two samples is that the treatment in the experimental sample is randomized, but the treatment in the observational sample is correlated with the pretreatment covariates that can be viewed as imperfect instruments. Table \ref{tab:data_requirement} illustrates the data requirements.

 \begin{table}[htbp!]
 \begin{threeparttable}
\caption{Data requirement: $\checkmark$ is observed} \label{tab:data_requirement}
\begin{tabular}{l|ccc|l}
\toprule
Dataset & Outcome & Covariates & Treatment  & Treatment Assignment\\
 & $Y$ & $Z$ & $X$ & \\
\midrule
Experimental & $\checkmark$ & $\checkmark$ & $\checkmark$ & $X$ is randomized \\
Observational & $\checkmark$ & $\checkmark$ & $\checkmark$ & $X$ is correlated with $Z$ \\
\bottomrule
\end{tabular}
\begin{tablenotes}
      \small
      \item Note: $Z$ is an imperfect instrument that may violate the exclusion restriction. 
    \end{tablenotes}
\end{threeparttable}
\end{table} 

I apply the method to estimate how online hotel rankings affect clicks using Expedia datasets in \cite{ursu_power_2018}.  The datasets include both an observational sample that ranks hotels based on relevance and an experimental sample that ranks hotels randomly.  In this context, the outcome of interest $Y$ is clicking on a given hotel,  the treatment $X$ is the ranking of that hotel,  and $Z$ includes rating and other hotel characteristics that may affect its ranking in the observational sample.  When compared to the estimate that only uses experimental data,  I show my method significantly improves the estimation efficiency by leveraging hotel characteristics as imperfect instruments. 

The remainder of the paper is organized as follows: Section 2 reviews the literature. Section 3 presents the setup and key assumptions. Section 4 introduces the estimation method. Section 5 discusses the theoretical efficiency gain.  Section 6 extends the method from linear to nonlinear models.  Section 7 illustrates the source and magnitude of efficiency gain through simulation. Section 8 applies the method to measure effects of hotel rankings on clicks using datasets from \cite{ursu_power_2018}. Section 9 concludes.

\section{Literature Review}

My paper is related to several contemporary studies that combine experimental and observational data. This strand of literature typically has two objectives: 1) identifying effects that cannot be identified using only experimental data, 2) improving the precision of estimates although experiment is sufficient for identification. Several papers focus on achieving the objective of identification. For example, \cite{athey_combining_2020} consider a case when the long-term treatment effect cannot be identified based only on the experimental data, because the long-term outcome in the experimental data is missing. \cite{kallus_removing_2018} consider a setting where the treatment effect for a certain population cannot be identified because the experiment does not cover this subpopulation. 

My paper focuses on improving efficiency instead of achieving identification. Even though experiments are often sufficient for identifying causal effects of interest, the estimate may be imprecise when large-scale experiments are costly or unrealistic. Given the limited experiment size, it is important to develop methods to improve the efficiency. \cite{rosenman2020combining} consider a case when the population is partitioned into several strata, and researchers are interested in improving average precision across strata. They show that the mean-squared-error can be improved by shrinking unbiased experimental estimates to additional biased observational estimates. However, when the number of strata is small or the bias is large, the efficiency gain could be small. \cite{peysakhovich_combining_2016} consider a case where researchers observe a panel of individuals that appear many times. They estimate individual-level observational estimates and use them as features to differentiate and sort individuals.\footnote{The method also requires a monotonic relationship between observational estimates and true causal effects. } Another popular method by \cite{deng_improving_2013} incorporates pre-experimental variables as additional covariates into the experimental analysis. Both \cite{deng_improving_2013} and \cite{peysakhovich_combining_2016} require observing individuals before and during the experiment, and is conceptually different from my method that is applicable even if individuals are only observed once. My method also has different determinants for efficiency gains. Method of \cite{deng_improving_2013} and \cite{peysakhovich_combining_2016} are most effective when the pre-experimental data contain variables that are highly predictive of the experimental outcome. My method is most effective when the pre-experimental data contain variables that are highly predictive of the endogenous treatment.
My method is complementary because it uses new assumptions that leverage a new source of information, allowing us to improve efficiency even when the number of strata is small or each individual is only observed once in the data. These methods are also not exclusive: they can be applied together if the data requirements and assumptions are met.

My estimation strategy is closely related to that of \cite{imbens_combining_1994}, who combine macro and micro data to improve estimation efficiency. My method is similar in that one dataset is sufficient for identifying the parameter of interest, and the moment conditions from the additional dataset improve efficiency. However, \cite{imbens_combining_1994} consider a setting in which all variables in both datasets are generated from the same distribution. In contrast, I consider a different setting in which the variable of interest is endogenously determined in the observational data but randomized in the experimental data.

My model for observational data is related to the literature by \cite{conley_plausibly_2010},  \cite{nevo_identification_2010}, and \cite{kippersluis_beyond_2018} that study plausibly exogenous models. In their models, some covariates are considered as ``plausible IV" or ``imperfect instruments" because they satisfy the relevance condition but may violate the exclusion restriction. \cite{conley_plausibly_2010} assume that there is a prior on how much the first-stage covariate violates the exclusion restriction. My method also depends on the availability of imperfect instruments that satisfy the relevance condition. However, my model allows the exclusion restriction to be strongly violated, and does not require a prior on the degree of violation, because this violation can be quantified by the experimental data.

\section{Setup}

For illustrative purposes, consider a linear causal model 
\begin{equation}\label{eqn:linear}
Y_i = \beta_1 X_i + \beta_2 Z_i + U_i \; ,
\end{equation}
where $Y_i$ is the outcome, $X_i$ is the focal variable of interest, $Z_i$ is the observed covariate, and $U_i$ is the unobserved covariate.  $\beta_1$ and $\beta_2$ are respectively the causal effect of $X_i$ and $Z_i$ on $Y_i$.  $\beta_1$ is the primary parameter of interest, and $\beta_2$ is a nuisance parameter that may or may not be 0.  

The main difference between the observational and the experimental data is how $X_i$ is generated. Let $G_i \in \{E, O\}$ be the indicator for the data a unit $i$ is drawn from, where $E$ denotes experimental data and $O$ denotes observational data. Assume the first-stage equations for $X_i$ to be:

\begin{equation}\label{eqn:x}
X_i = \begin{cases} 
\gamma Z_i + V_i & \text{ if }G_i = O\\
randomized & \text{ if } G_i = E\\
\end{cases}. 
\end{equation}
$\gamma$ is a non-zero first-stage coefficient for the covariate $Z_i$ in the observational data and $V_i$ is the corresponding unobserved error term.\footnote{The first-stage equation does not have to be linear nor causal. $Z_i$ can be thought of as a composite score of consumer characteristics after applying nonlinear rules on multiple covariates, and this nonlinear rule can be estimated using any flexible predictive models. Following the two-stage-least-square method, the coefficient $\gamma$ can also be understood as the best linear predictor for $X$ given $Z$. } Assume $X_i$ in the observational data is correlated with the unobserved covariate $U_i$ so that the observational data has an endogeneity problem:

$$
Cov(X_i, U_i|G_i = O) \neq 0 \; .
$$
Also assume $Z_i$ violates the exclusion restriction so that it is not a valid instrumental variable (IV). This violation can be caused by a direct effect on $Y_i$ or a correlation with the unobserved covariate:
$$
\beta_2 \neq 0 \; \text{ or }\; Cov(Z_i, U_i|G_i = O) \neq 0 \; .
$$
Compared to the observational data, the experimental data can identify $\beta_1$ because $X_i$ is randomized and thus independent of $U_i$:
$$
X_i \perp U_i |G_i = E \; .
$$
To summarize the basic setup, the experimental data are sufficient to identify $\beta_1$, but the accuracy is limited by the sample size.  The observational data alone cannot identify $\beta_1$ because $X_i$ is endogenous and $Z_i$ is an imperfect instrument - a covariate that satisfies the first-stage relevance condition but violates the exclusion restriction.  The goal of this paper is to leverage this imperfect instrument $Z_i$ to improve estimation efficiency.

\subsection{Assumptions}

The key assumption is that the joint distributions of $(Z_i, U_i)$ are similar between experimental and observational datasets, such that information learned in one dataset can be transferred to another dataset. One sufficient condition is that the distribution is identical between experimental and observational data: 

\let\origtheassumption\theassumption
\edef\oldassumption{\the\numexpr\value{assumption}+1}

\begin{assumption}[Identical Distribution of Covariates]\label{asmp:random_sampling}
$$
G_i \perp (Z_i, U_i) \; ,
$$
\end{assumption}
Assumption \ref{asmp:random_sampling} holds if a unit $i$ with attributes $(Z_i, U_i)$ is first randomly sampled from the population and then randomly assigned into the experimental or observational group, independent of $(Z_i, U_i)$. Figure \ref{fig:sampling_procedure} illustrates an experimental procedure in the digital environment that satisfies this assumption:
\begin{figure}[H]
\centering
\includegraphics[scale=0.5]{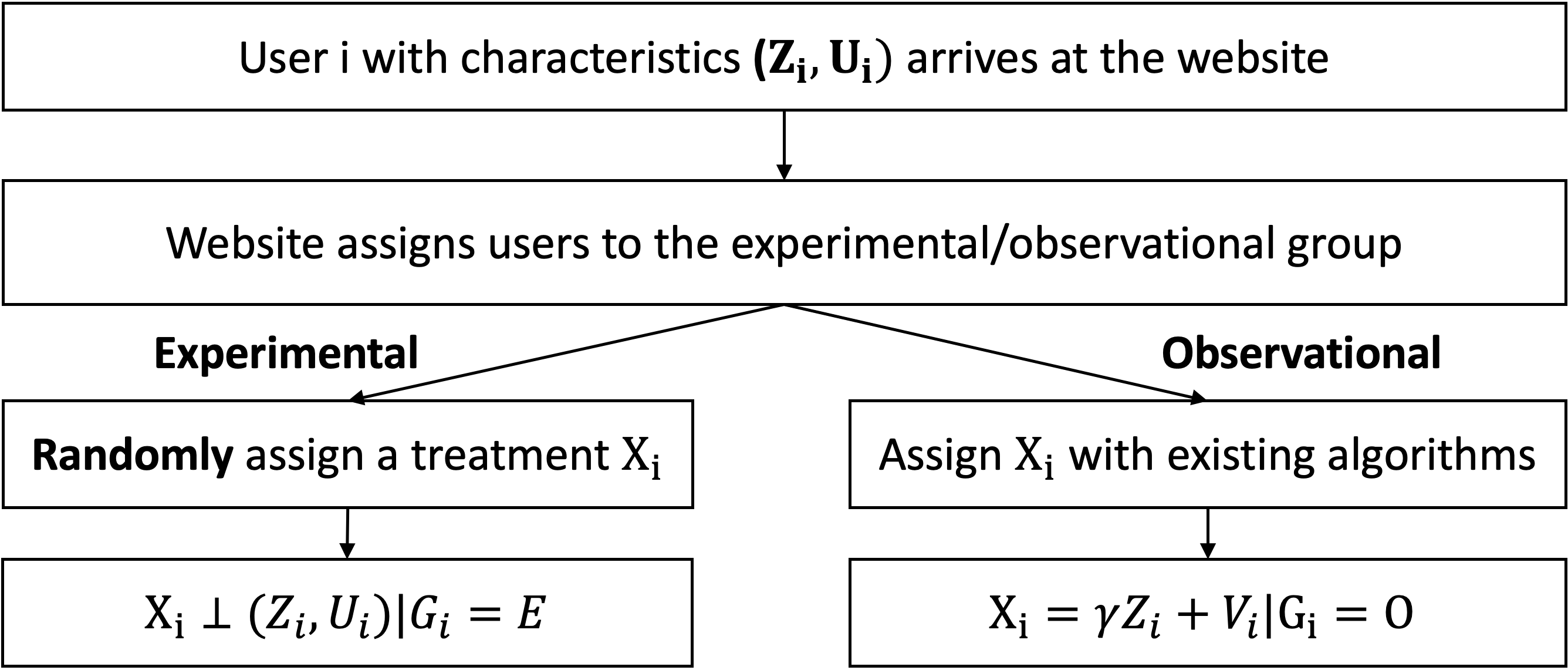}
\caption{An experimental design example that randomly assigns users to the experimental/observational group}\label{fig:sampling_procedure}
\end{figure}

This experimental  procedure is commonly used for experiments that shuffle the rankings of search results or ads displayed to users. Google \citep{wang_learning_2016}\footnote{As described in the randomization section of the paper: given a ranked result list of n documents returned for some query, instead of showing the original list, we permute the results uniformly at random and present the shuffled list to a small fraction of end users} and Microsoft  \citep{li_counterfactual_2015} have conducted experiments that first randomly select a small fraction of users, and then randomize the search results displayed to those users. Expedia \citep{ursu_power_2018} has also run such an experiment that shuffles the ranks of hotels displayed to a subset of customers. $JD.com$ \citep{carrion_blending_2021} has conducted experiments that randomly shuffle the orders of ads for a random subset of users.  In these scenarios, $Z$ can be the relevance score of a website, $X$ is the rank of a website, and $Y$ is the outcome of interest such as a click. By conducting these experiments, firms can answer counterfactual questions such as how likely a customer is going to click a certain webpage when it has a ranking of $X$, a crucial input for firms to improve ranking and user engagement.

Another possibility is that the observational sample includes customers who arrived in the past, while the experimental sample includes customers who arrived more recently.  If past customers differ significantly from future customers, firms will have limited incentive to conduct experiments, because findings from past customers have limited value for future strategies.  Therefore, for firms that do have incentive to conduct experiments,  it is plausible to assume that past and future customers are similar.  Given this assumption,  combining past observational data and current experimental data can improve efficiency. 

The remainder of the paper discusses the estimation strategy that combines datasets that have similar distributions in covariates $(Z_i, U_i)$.  For illustrative purposes,  I first focus on the estimation strategy for the simple case when the model is linear and the distribution of $(Z_i, U_i)$ is identical.  These assumptions allow me to better articulate the intuition of my method as well as analytically quantifying the magnitude of the efficiency gain.  Section \ref{sec:extension_nonlinear} and Appendix B.3 
 discuss how the method can be extended to a class of nonlinear models.  Appendix B.1 
discusses how Assumption \ref{asmp:random_sampling} can be relaxed to allow for the distribution to be similar but non-identical.   

\section{Estimation Strategies}

\subsection{Intuition}
To develop intuition, consider when $Z_i$ is used as an imperfect IV to estimate $\beta_1$ in observational data. Define the probability limit of this estimator as $b_1^{IV}$: 
\begin{equation}\label{eqn:biv}
b_{1}^{IV} \equiv \frac{Cov(Y_i, Z_i|G_i = O)}{Cov(X_i, Z_i|G_i = O)}.  
\end{equation}

This estimator is biased and inconsistent because $Z_i$ violates the exclusion restriction, either due to a direct effect on $Y_i$ or an unobserved correlation with $U_i$:
\begin{equation}\label{eqn:proof_bias}
\begin{aligned}
b_1^{IV} & = \frac{Cov(Y_i, Z_i|G_i = O)}{Cov(X_i, Z_i|G_i = O)}  \\
& = \frac{Cov(\beta_1 X_i + \beta_2 Z_i + U_i, Z_i|G_i = O)}{Cov(X_i, Z_i|G_i = O)}  \\
& = \beta_1 + \frac{\beta_2 Var(Z_i|G_i = O) + Cov(U_i, Z_i|G_i = O)}{Cov(X_i, Z_i|G_i = O)}\\
& = \beta_1 + \frac{Var(Z_i|G_i = O)}{Cov(X_i, Z_i|G_i = O) }\frac{\beta_2 Var(Z_i|G_i = O) + Cov(U_i, Z_i|G_i = O)}{Var(Z_i|G_i = O)}\\
& = \beta_1 + \frac{1}{\gamma} \left(\underbrace{\beta_2}_{\text{Direct effect of Z}} + \underbrace{\frac{Cov(U_i, Z_i|G_i = O)}{Var(Z_i|G_i = O)}}_{\text{Unobserved correlation}}\right).
\end{aligned}
\end{equation}
Let $b_2$ denote the violation of such exclusion restriction: 
\begin{equation}\label{eqn:b2}
b_2 \equiv \beta_2 + \frac{Cov(U_i, Z_i|G_i = O)}{Var(Z_i|G_i = O)}.
\end{equation}
A direct implication of Equation \ref{eqn:proof_bias} and \ref{eqn:b2} is:
\begin{lemma}\label{lemma:bias}
The bias of the imperfect IV estimator is $\frac{b_2}{\gamma}$. 
\end{lemma}

When using observational data alone, this bias typically cannot be corrected because the violation of exclusion restriction $b_2$ is unknown and cannot be estimated. However, approximating this violation becomes possible with experimental data by regressing $Y_i$ on $Z_i$. Define the probability limit of this estimated coefficient as $b_2^E$:
$$
b_2^E \equiv \frac{Cov(Y_i, Z_i|G_i = E)}{Var(Z_i|G_i = E)}.
$$
\begin{lemma}\label{lemma:violation}
Under Assumption \ref{asmp:random_sampling}, the violation of exclusion restriction $b_2$ is identified by estimating $b_2^E$ using experimental data.
\end{lemma}

\noindent \textbf{Proof}:
$$
\begin{aligned}
b_2^E  & \equiv \frac{Cov(Y_i, Z_i|G_i = E)}{Var(Z_i|G_i = E)}\\
& = \frac{Cov(\beta_1 X_i + \beta_2 Z_i + U_i, Z_i|G_i = E) }{Var(Z_i|G_i = E)}\\
& = 0 + \beta_2 + \frac{Cov(U_i, Z_i|G_i = E)}{Var(Z_i|G_i = E)}\\
& = \beta_2 + \frac{Cov(U_i, Z_i|G_i = O)}{Var(Z_i|G_i = O)} \\ 
& = b_2
\end{aligned}
$$

To summarize the intuition, the imperfect IV estimate in the observational data is more useful when its bias can be quantified, and the experimental data help quantify such bias. Next I propose two different estimation procedures,  weighting and GMM.  These two procedures are qualitatively the same, but give different intuitions to help understand the source of efficiency gain.

\subsection{Weighting}\label{sec:weighting}

\begin{figure}[H]
\centering
\includegraphics[scale=0.5]{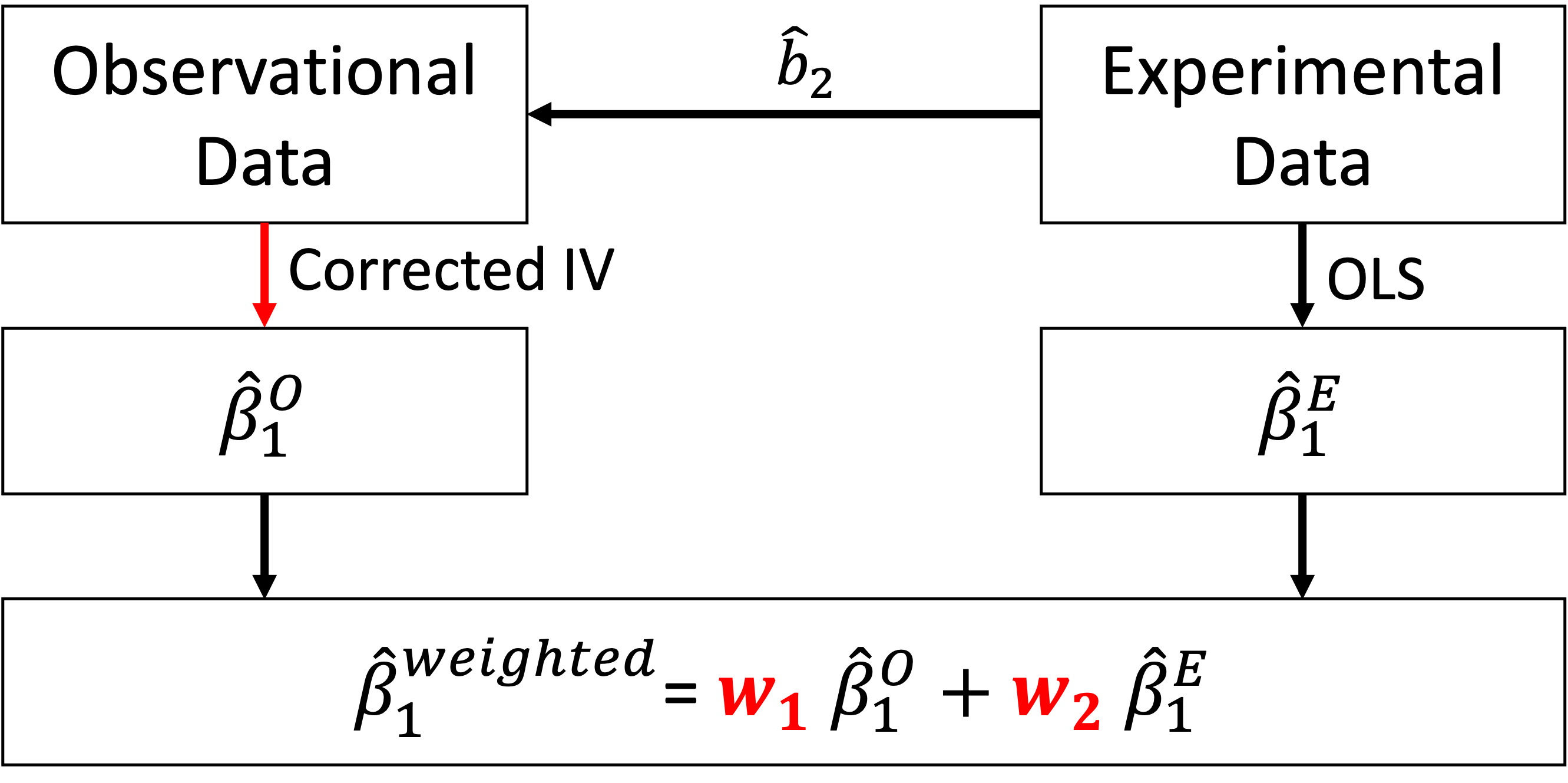}
\caption{Combining observational and experimental data through weighting}\label{fig:weighting_procedure}
\end{figure}

One procedure is to derive two consistent estimators of $\beta_1$ that are uncorrelated and then combine them through weighting. This procedure involves multiple steps visualized in Figure \ref{fig:weighting_procedure}:
 
\begin{enumerate}
\item  Regress $Y_i$ on ($X_i$, $Z_i$) in experimental data to obtain $(\widehat{\beta_1}^E, \widehat{b_2}^E)$.\footnote{Note that even though experimental data are insufficient to identify $\beta_2$ due to the potential correlation between $U$ and $Z$, $b_2$ can be identified. So the method focuses on $(\widehat{\beta_1}^E, \widehat{b_2}^E)$ instead of $(\widehat{\beta_1}^E, \widehat{\beta_2}^E)$}
\item Use $Z_i$ as an imperfect $IV$ in observational data to obtain $\widehat{b_1}^{IV}$.
\item Regress $X_i$ on $Z_i$ in observational data to obtain $\widehat{\gamma}^O$.
\item Use Lemma \ref{lemma:bias} to correct the bias of the imperfect IV estimator:  $\widehat{\beta_1}^O  = \widehat{b_1}^{IV} - ({\widehat{b_2}^E}/{\widehat{\gamma}^O})$.
\item Take a weighted average of the two estimates of $\widehat{\beta_1}^{weighted} = w_O \widehat{\beta_1}^O + w_E\widehat{\beta_1}^E$, where the weight $(w_O, w_E)$ are hyperparameters chosen to minimize $Var(\widehat{\beta_1}^{weighted})$.
\end{enumerate} 
 
\begin{lemma}\label{lemma:uncorrelation}
$\widehat{\beta_1}^O$ and $\widehat{\beta_1}^E$ are uncorrelated.
\end{lemma}
\begin{lemma}\label{lemma:inverse_weighting}
Under Lemma \ref{lemma:uncorrelation} the weighting minimizes the variance of the combined estimator, if each estimate is weighted in inverse proportion to its variance:
$$
\frac{w_O^*}{w_E^*} = \frac{Var(\widehat{\beta_1}^E)}{Var(\widehat{\beta_1}^O)}
$$
\end{lemma}
I prove Lemma \ref{lemma:uncorrelation} and Lemma \ref{lemma:inverse_weighting} in Appendix \ref{proof:uncorrelation}. 

This weighting method helps explain the determinants of efficiency gain, which depends on the accuracy of $\widehat{\beta_1}^{O} = \widehat{b_1}^{IV} - \left(\frac{\widehat{b_2}^E}{\widehat{\gamma}^O}\right)$. The accuracy of $(\widehat{b_1}^{IV}, \widehat{\gamma}^O)$ is high if the observational sample is large. When the observational sample tends to infinity such that $(\widehat{b_1}^{IV}, \widehat{\gamma}^O)$ become exact estimates of $(b_1^{IV}, \gamma)$, $\widehat{\beta_1}^{O}$ becomes an unbiased estimate whose only source of uncertainty comes from $Var\left(\frac{\widehat{b_2}^E}{\gamma}\right)$. This remaining uncertainty would be small if the first-stage relevance parameter $\gamma$ is large, or if the variation of $Z_i$ is large such that $Var(\widehat{b_2}^E)$ is small. To summarize, the overall efficiency gain is large if the observational sample is large and includes relevant pretreatment covariates $Z_i$ that can explain variation in treatment assignment $X_i$.

\subsection{GMM}

I also consider a GMM approach, which enables practitioners to easily derive the estimator's analytical solution and its asymptotic properties using established GMM machinery. Equation \ref{eqn:linear} is not suitable for direct GMM estimation however, because one of the parameters $\beta_2$ cannot be identified when $Cov(U_i, Z_i) \neq 0$. To derive relevant moment conditions, it is convenient to work with a residual $\epsilon_i$ that is determined by identifiable parameters $(\beta_1, b_2)$:
\begin{equation}\label{eqn:epsilon}
\epsilon_i \equiv Y_i - \beta_1 X_i - b_2 Z_i.
\end{equation}  

\begin{lemma}\label{lemma:epsilon}
If Assumption \ref{asmp:random_sampling} is satisfied, then 
$$
\epsilon_i =  U_i - \frac{Cov(U_i, Z_i)}{Var(Z_i)}Z_i 
$$
such that $\epsilon_i$ and $Z_i$ are not correlated in both the observational and experimental data:
$$
Cov(\epsilon_i, Z_i|G_i = g) = 0 \; \text{for $g \in \{O, E\}.$}
$$ 
\end{lemma}
I prove Lemma \ref{lemma:epsilon} in Appendix \ref{proof:epsilon}. Note $\epsilon_i$ is different from $U_i$ and can be understood as a transformation of $U_i$ that removes its correlation with the observed covariate $Z_i$.  For illustrative purpose, consider the simple case when $(Y_i, X_i, Z_i)$ are already demeaned based on Frisch-Waugh-Lovell theorem such that $E[\epsilon_i] = 0$.\footnote{The idea can be easily extended to cases when the model has an intercept and its variables have non-zero means based on Frisch-Waugh-Lovell theorem.  Consider any variables $(\tilde{Y}_i, \tilde{X}_i, \tilde{Z}_i)$ that satisfy $\tilde{Y} = \alpha + \beta_1 \tilde X_i + b_2 \tilde Z_i + \epsilon_i$,  this model can be reduced to Equation \ref{eqn:epsilon} by defining $(Y, X, Z)$ as demeaned residuals, where $Y_i = \tilde{Y}_i - E[\tilde{Y}_j|G_j = G_i]$,  $X_i = \tilde{X}_i - E[\tilde{X}_j|G_j = G_i]$, $Z_i = \tilde{Z}_i - E[\tilde{Z}_i]$.} Based on Lemma \ref{lemma:epsilon}, two moment conditions can be derived from the experimental data
$$
\begin{aligned}
E[(Y_i - \beta_1 X_i  - b_2 Z_i)X_i|G_i = E] = 0\\
E[(Y_i - \beta_1 X_i  - b_2 Z_i)Z_i|G_i = E] = 0\\
\end{aligned}
$$
and one extra moment condition from the observational data:
$$
\begin{aligned}
E[(Y_i - \beta_1 X_i  - b_2 Z_i)Z_i|G_i = O] = 0.
\end{aligned}
$$
Denote the vector of moment functions as $g$: 

$$
g_i(\beta_1, b_2) = 
\begin{bmatrix}
(Y_i - \beta_1 X_i - b_2 Z_i ) X_i I(G_i = E) \\
(Y_i - \beta_1 X_i - b_2 Z_i ) Z_i I(G_i = E)\\
 (Y_i - \beta_1 X_i - b_2 Z_i ) Z_i I(G_i = O)
\end{bmatrix}
$$
and the GMM estimator can be written as 
\begin{equation}\label{eqn:gmm_minimization}
(\widehat{\beta_1}^{GMM}, \widehat{b_2}^{GMM}) = \text{argmin} [\frac{1}{N} \sum_{i = 1}^N g_i(\widehat{\beta_1}, \widehat{b_2}) ]' \cdot W \cdot [\frac{1}{N} \sum_{i = 1}^N g_i(\widehat{\beta_1}, \widehat{b_2}) ]
\end{equation}
where $W$ is a non-negative definite weighting matrix.  The optimal $W^*$ depends on the covariance matrix of the moment conditions, which can be estimated by running the standard feasible GMM.  Appendix \ref{appendix:gmm_analytical} derives the analytical solution of this GMM estimator. 

\section{Efficiency Gain}  \label{sec:efficiency_gain}
 
In this section I illustrate the efficiency gain using the GMM approach.  It is convenient to first formulate the estimation problem in matrix notation:
\begin{equation}\label{eqn:reformulation}
\mathbf{Y} = \beta_1 \mathbf{X} + b_2 \mathbf{Z} + \mathbf{\epsilon}
\end{equation}
where each row is
$$
Y_i = \beta_1 X_i + b_2 Z_i + \epsilon_i,
$$
and the unit is ordered by group $g \in \{E, O\}$ such that 
$$
[\mathbf{Y},  \mathbf{X}, \mathbf{Z}, \mathbf{\epsilon}] = 
\begin{bmatrix}
\mathbf{Y}_E & \mathbf{X}_E & \mathbf{Z}_E & \epsilon_E\\
\mathbf{Y}_O & \mathbf{X}_O & \mathbf{Z}_O & \epsilon_O
\end{bmatrix}
$$
Since the exact efficiency gain depends on the distribution of $(X_i, Z_i, \epsilon_i)$, I focus on the simple case of homoscedasticity and nonautocorrelation
\begin{assumption}{(Homoscedasticity and Nonautocorrelation)}\label{asmp:no_homo_auto}
$$
E[\epsilon \epsilon' |\mathbf{X_E}, \mathbf{Z}, \mathbf{G}] = \Sigma = \sigma^2 I,
$$
\end{assumption}
This assumption is not essential for the method to improve efficiency but helps illustrate the determinants of efficiency gain.\footnote{$\mathbf{X_O}$ is also not conditioned on because it is an endogenous variable.}

\begin{theorem}\label{thm:basic} 

If Assumptions \ref{asmp:random_sampling} and \ref{asmp:no_homo_auto} are satisfied, an efficient GMM estimator has an asymptotic variance of:
$$
\mathbb{V}(\widehat{\beta_1}^{GMM}) = \frac{\sigma^2}{n_E} \left[Var(X_i|G_i = E) + \pi_O Var(\gamma Z_i|G_i = O)\right]^{-1}.
$$
\end{theorem}
The proof for this theorem is provided in Appendix \ref{proof:theorem_gmm}.  In comparison, the asymptotic variance of the experiment-only estimator is:
$$
\mathbb{V}(\widehat{\beta_1}^{E}) =  \frac{\sigma^2}{n_E} \left[Var(X_i|G_i = E) \right]^{-1}
$$
A comparison of these two variances suggests that the efficiency gain is determined by 1) the proportion of observational data, 2) the relevance of $Z_i$ as an imperfect instrument, and 3) the relative variance of $\mathbf{X}_E$ and $\mathbf{X}_O$:

\begin{lemma} \label{lemma:same_v_gain}
If $Var(X_i|G_i = O) = Var(X_i|G_i = E)$,\footnote{
This equality assumption is easily satisfied in ranking experiments in online settings, where positions of ads or hotels listed by a website are randomly shuffled. It is also satisfied if the experimental $X$ is sampled from the marginal distribution of the observational $X$. 
} the relative variance of the two estimators can be written as:
\begin{align*}
 \frac{\mathbb{V}(\hat{\beta}^{E}_1) }{\mathbb{V}(\hat{\beta}^{GMM}_1 )} & = 1 + \pi_{O} \times \frac{Var(\gamma Z_i|G_i = O)}{Var(X_i|G_i = O)} \times \frac{Var(X_i|G_i = O)}{Var(X_i|G_i = E)}\\
 & = 1 + \pi_O \times R_{x, z, O}^2
\end{align*}
where  $\pi_O$ is the proportion of observational data,  and $R_{x, z, O}^2$ is the first-stage R-squared that summarizes how well the variation in $Z$ explains the variation in $X$ in the observational data.
\end{lemma}

Given that $R_{x, z, O}^2 \in [0, 1]$ and  $\pi_O \in [0,  1)$,  Lemma \ref{lemma:same_v_gain} implies $ \frac{\mathbb{V}(\hat{\beta}^{E}_1) }{\mathbb{V}(\hat{\beta}^{GMM}_1 )} \in [1, 2)$, which means that the method can reduce the variance by up to $50\%$.  This potential improvement implies that half of the experimental data are required to achieve the same accuracy when observational data are incorporated.

\section{Extensions to Nonlinear Model}\label{sec:extension_nonlinear}

The paper so far has focused on simple linear models specified in Equation \ref{eqn:linear}.  This section considers how the method can be generalized to a class of nonlinear models:

\begin{equation}\label{eqn:parametric_nonlinear}
Y_i = f(X_i, Z_i; \tilde{\theta}) + U_i
\end{equation}
where $f$ is a flexible nonlinear function parameterized by $\tilde{\theta}$.  Analogous to the linear case,  $\tilde{\theta}$ may not be directly identified by the experimental data when $Cov(U_i, Z_i) \neq 0$.  It is thus convenient to work with a transformed function $h$: 
\begin{equation}\label{eqn:nonlinear_transformation}
h(X_i, Z_i; \theta) \equiv f(X_i, Z_i; \tilde{\theta}) + E[U_i|Z_i, G_i = E]
\end{equation}
where $\theta$ are the unknown parameters to be estimated.   

The function $h(X, Z; \theta)$ can be understood as the conditional average potential outcome for units with attributes $Z$ that receives treatment $X$. Accurately estimating this function is valuable because the function can be used to derive the average treatment effect (ATE), the conditional average treatment effect (CATE), as well as the marginal effect.\footnote{For example, when $X$ is binary, $h(1, Z; \theta) - h(0, Z; \theta)$ is CATE conditional on a specific $Z$, and $E[h(1, Z_i; \theta) - h(0, Z_i; \theta)|G_i = E]$ is the ATE for the experimental population. When $X$ is continuous, $\frac{\partial h(X, Z; \theta)}{\partial X}$ is the marginal effect. } When researchers are interested in effects such as CATE and marginal effects instead of ATE, it becomes even more important to improve precision because the variances of CATE and marginal effects are typically larger than the variance of ATE. 

Recall that I focus on cases when experimental data are sufficient for identification.  Suppose that the experimental data estimate $\theta$ by minimizing an objective function $L_E(\mathbf{Y}_E, \mathbf{X}_E, \mathbf{Z}_E; \theta)$
\begin{equation}
\widehat{\theta}^E = arg\min_{\theta} L_E(\mathbf{Y}_E, \mathbf{X}_E, \mathbf{Z}_E; \theta).
\end{equation}
The objective function $L_E$ can take different forms.  It can be the squared loss function $\sum_{i:G_i = E}(Y_i - h(X_i, Z_i; \theta))^2$, based on the nonlinear least square method. Alternatively, it can be a (negative) likelihood function based on distributional assumptions about the error term, using the maximum likelihood approach. It can also be an objective function based on a set of moment conditions using GMM. 

Analogous to the linear case,  observational data help improve efficiency because Assumption \ref{asmp:random_sampling} implies that another moment condition can be derived from the observational data:\footnote{Assumption  \ref{asmp:random_sampling} implies $E[U_i|Z_i, G_i = E] = E[U_i|Z_i, G_i = O]$,  which means $E[(Y_i - h(X_i, Z_i; \theta)) Z_i|G_i = O] =  E[(U_i - E[U_i|Z_i, G_i = E]) Z_i|G_i = O] = E[(U_i - E[U_i|Z_i, G_i = O])Z_i|G_i = O] = 0$. }
$$
E[(Y_i - h(X_i, Z_i; \theta)) Z_i|G_i = O] = \mathbf{0}
$$
A sample analog of this moment condition is
$$
m(\mathbf{Y}_O, \mathbf{X}_O, \mathbf{Z}_O; \theta) = \frac{1}{N_O}\sum_{i:G_i = O}(Y_i - h(X_i, Z_i; \theta))Z_i
$$
If the observational data are sufficiently large, then $m$ should be close to $0$ when $\widehat{\theta}$ is accurate.  Any deviation from $0$ should be penalized because it implies that $\widehat{\theta}$ is inaccurate.  For univariate $Z_i$, this penalization can be incorporated by adding an additional regularization term into the original objective function:
\begin{equation}
\widehat{\theta}^{combine} = arg\min_{\theta} L_E(\mathbf{Y}_E, \mathbf{X}_E, \mathbf{Z}_E; \theta) + \lambda m(\mathbf{Y}_O, \mathbf{X}_O, \mathbf{Z}_O; \theta)^2
\end{equation}
where $\lambda$ is a non-negative hyperparameter that determines how much to penalize the deviation from the moment condition. Intuitively, $\lambda$ depends on how much researchers believe the violation of such constraint can be attributed to the inaccuracy of $\widehat{\theta}$. 

When $\lambda = 0$, it implies that researchers ignore information from observational data, and the optimization problem is equivalent to the one using only experimental data. A small value of $\lambda$ can be rationalized if the observational dataset is small, such that the violation of additional moment condition can be explained by sampling error rather than the inaccuracy of $\theta$. 

When $\lambda = \infty$, it implies that researchers heavily rely on the constraint. A large $\lambda$ can be justified if the observational dataset is large such that any violation of the moment condition must be attributed to the inaccuracy of $\theta$.  When $\lambda = \infty$, the problem also becomes a constraint optimization that helps improve efficiency (\cite{imbens_combining_1994}). 

For multivariate $Z_i$ and a vector of moment conditions $m$, the framework can be extended by taking a weighted average of these moment conditions and minimizing:\footnote{$Z_i$ can also include the intercept, which implies a moment condition $E[(Y_i - h(X_i, Z_i;\theta)1|G_i = O] = 0$.}
\begin{equation}
\widehat{\theta}^{combine} = arg\min_{\theta} L_E(\mathbf{Y}_E, \mathbf{X}_E, \mathbf{Z}_E; \theta) + \lambda m(\mathbf{Y}_O, \mathbf{X}_O, \mathbf{Z}_O; \theta)'W_O m(\mathbf{Y}_O, \mathbf{X}_O, \mathbf{Z}_O; \theta)
\end{equation}
where $W_O$ is a weighting matrix that determines how much weight to choose for each moment.  The choice of $W_O$ only affects the efficiency improvement and does not affect identification,  because experimental data are already sufficient for identification.  One possible option of $W_O$ is $(\mathbf{Z}_O'\mathbf{Z}_O)^{-1}$,  which mimic the weight in the linear GMM case.

The exact efficiency gain of incorporating these additional moment conditions depends on the functional format of $h$,  what types of variables are available, and what the objectives are. Providing an analytical formula for this more general case is beyond the scope of this paper.  Section \ref{sec:empirical} illustrates how to apply this method in the empirical setting of online ranking and Appendix B.2 
 further discusses the intuition for efficiency gain from a GMM perspective.  The key insight is that the local moment conditions of these nonlinear cases mimic the moment conditions in the linear cases, such that similar efficiency gain could be achieved.

One limitation of this class of non-linear models is that the unobserved $U_i$ is additively separable from the observed $(X_i, Z_i)$.  When $U_i$ is not additively separable, it is well known that the endogeneity is difficult to handle and has only been extensively studied in a few special cases.  Appendix B.3 
discusses how a similar method can still be applied in binary probit model, of which the error term is not \textit{directly} separable.  

\section{Simulation}

This section uses simulation to illustrate the effectiveness of the method and the source of efficiency gain.  Consider a linear causal model 
\begin{equation}\label{eqn:simulation_basic}
Y_i = \beta_1 X_i + \beta_2 Z_i + U_i
\end{equation}
where $X_i$ is endogenously determined by $(Z_i, V_i)$ in the observational group but randomly assigned in the experimental group:
$$
X_i = \begin{cases} 
\gamma Z_i + V_i & \text{ if }G_i = O\\
\text{drawn from }N(0, 1) \text{ randomly} & \text{ if } G_i = E\\
\end{cases} 
$$
Unit $i$ with characteristics $(Z_i, U_i, V_i)$ is drawn from a distribution unknown to researchers:
$$
(Z_i, U_i, V_i) \sim N \left(0, 
\begin{bmatrix}
1 & 0.4\sigma_u & 0\\
0.4\sigma_u & \sigma_u^2 & 0.4 \sigma_u \sigma_v\\
0 & 0.4 \sigma_u \sigma_v & \sigma_v^2
\end{bmatrix}
\right)
$$
$V_i$ is the residual term in the observational first-stage and therefore by definition uncorrelated with $Z_i$.  I calibrate $\sigma_v$ and $\gamma$ such that $V(X_i|G_i = O) = 1 =  V(X_i|G_i = E)$ and $R^2_{X, Z, O} = 0.6$,  and I assume that the dataset has a small number of experimental units with $n_E = 100$ and a large number of observational units with $n_O = 1900$.

To make the simulation concrete,  consider an advertiser that is interested in measuring the return to advertising (\cite{lewis_unfavorable_2015}). Let $Y_i$ denote the net profit of advertising to customer $i$,  $X_i$ denote the ad expenditure on customer $i$, and $(Z_i, U_i)$ be the observed and unobserved customer characteristics that the advertiser uses to target customers and determine expenditure.  After certain observational periods, the advertiser becomes interested in measuring the advertising ROI.  The advertiser realizes that estimates obtained using past observational data may be biased due to endogeneity and decide to run an experiment, and make future advertising decisions based on the ROI estimated using this experiment. If the ROI estimate is not significantly higher than a certain threshold, the advertiser may choose to stop advertising in the future. Therefore, whether the ROI can be accurately estimated is crucial for advertiser's future decisions.  Unfortunately, when the advertising effect is small, it is difficult to precisely estimate advertising ROI.  For example, \cite{lewis_unfavorable_2015} conducted several large-scale experiments with major U.S. retailers and documented that advertisers may not be able to distinguish a campaign with high ROI of $50\%$ from a campaign with $0\%$ ROI because the median standard error on ROI for their experiment is $26.1\%$.

To simulate such a scenario,  I calibrate the simulation exercise such that the ROI is $\beta_1 = 50\%$ and the expected standard error of this ROI is $sd(\beta_1) = 26.1\%$ if it is estimated only using experimental data.\footnote{$sd(\beta_1)$ can be calibrated by adjusting $\sigma_u^2$.} To evaluate the performance of the proposed method,  I draw $10,000$ samples,  where each sample contains $n_E$ experimental units and $n_O$ observational units.  Within each sample $s$, I calculate the experiment-only estimate $\widehat{\beta_1}^{E,s}$ and the combined GMM estimate $\widehat{\beta_1}^{Combine,s}$. Table \ref{tab:simulation_example} gives an example of the regression results estimated using one such sample, where the combined GMM estimator detects that the coefficient is significantly different from 0. 

\begin{table}[!htbp]
  \caption{A sample where combined GMM estimator improves statistical significance} 
    \centering\small
  \label{tab:simulation_example} 
\begin{tabular}{@{\extracolsep{5pt}}lcc} 
\\[-1.8ex]\hline 
\hline \\[-1.8ex] 
 & \multicolumn{2}{c}{\textit{Dependent variable: Y}} \\ 
\cline{2-3} 
\\[-1.8ex] Estimator & Experiment-Only & Combined\\ 
\hline \\[-1.8ex] 
 X & 0.215 & 0.429$^{**}$ \\ 
  & (0.284) & (0.211) \\ 
\hline \\[-1.8ex] 
Experiment Data & 100 & 100 \\ 
Observational Data & & 1900\\
Ground Truth & 0.5 & 0.5\\
\hline 
\hline \\[-1.8ex] 
\textit{Note:}  & \multicolumn{2}{r}{$^{*}$p$<$0.1; $^{**}$p$<$0.05; $^{***}$p$<$0.01} \\ 
\end{tabular} 
\end{table} 

\begin{figure}[htbp!]
\centering
\includegraphics[scale=0.7]{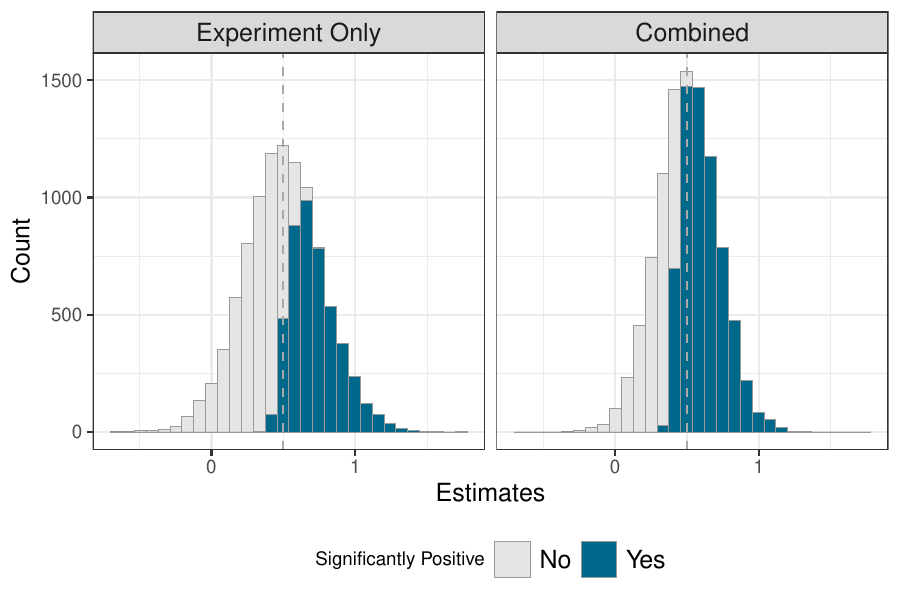}
\caption{Distribution of Estimates Across 10,000 Samples}\label{fig:simulation_dist_comparison}
\small \textit{Notes}: vertical line represents the true value of $\beta_1$ 
\end{figure}

After performing the estimation for each sample, I then summarize the bias, variance, and MSE across all samples. Figure \ref{fig:simulation_dist_comparison} compares the distributions of coefficients estimated by these two methods and shows that the combined method is more accurate.

Table \ref{tab:benchmark_summary_simulation} compares the combined GMM estimator with the observational methods including OLS and IV.\footnote{Experimental estimated is obtained by regressing $Y_E$ on $X_E$ and $Z_E$; observational OLS  is obtained by regression $Y_O$ on $X_O$ and $Z_O$; IV is obtained by using $Z_O$ as an instrument for $X_O$. } The last column of Table \ref{tab:benchmark_summary_simulation} documents the MSE of these estimators relative to that of the experiment-only estimator. The combined GMM estimator reduces the MSE by almost $1 - 0.607 = 39.3\%$.  In contrast,  methods based only on observational data have high MSE due to high bias: OLS is biased because $X$ is highly endogenous, and IV is biased because $Z$ is not a valid IV. 

\begin{table}[htbp!]
\caption{Summary statistics for the performance of estimators over $10,000$ samples}
\label{tab:benchmark_summary_simulation}

\begin{tabular}{lllll}
\toprule
Method & Bias$^2$ & Variance & MSE & Relative MSE\\
\midrule
Experiment Only & 0.000 & 0.074 & 0.074 & 1.000\\
Combined GMM & 0.000 & 0.045 & 0.045 & 0.607\\
OLS (Obs) & 1.624 & 0.007 & 1.632 & 21.936\\
IV (Obs) & 1.366 & 0.005 & 1.371 & 18.431\\
\bottomrule
\end{tabular}

\end{table}

To illustrate the implications of the improved efficiency,  Table \ref{tab:benchmark_summary_significance} compares how often the estimated coefficient has the correct sign (positive), and how often it is statistically significant. Both experimental and combined estimators generate the correct sign for most of the samples. However, compared to the experiment-only estimator, the combined estimator is $64.97\% - 45.24\% =  19.73\%$ more likely to detect that the effect is significant.  Suppose the advertiser makes future advertising decisions based on statistical significance,  then the advertiser is $19.73\%$ more likely to make the correct decision of keep running this profitable campaign by combining datasets.\footnote{Although it is common for firms to make decisions based on statistical significance,  \cite{feit_test_2019} highlight that it may not be optimal to do so and derive an optimal profit-maximizing test size.  Since my method reduces variance,  it can also improve profit by further reducing the profit-maximizing test size }

\begin{table}[htbp!]
\caption{Summary of statistical significance  ($p < 0.05$) over $10,000$ samples}
\label{tab:benchmark_summary_significance}

\begin{tabular}{lll}
\toprule
 & Positive (Correct Sign) & Significantly Positive\\
\midrule
Experiment Only & 96.56\% & 45.24\%\\
Combined GMM & 99.00\% & 64.97\%\\
\bottomrule
\end{tabular}

\end{table}

\subsection{Determinants of Efficiency Gain}

\begin{figure}[H]
\begin{subfigure}{.48\textwidth}
  \centering
  \includegraphics[width=.95\linewidth]{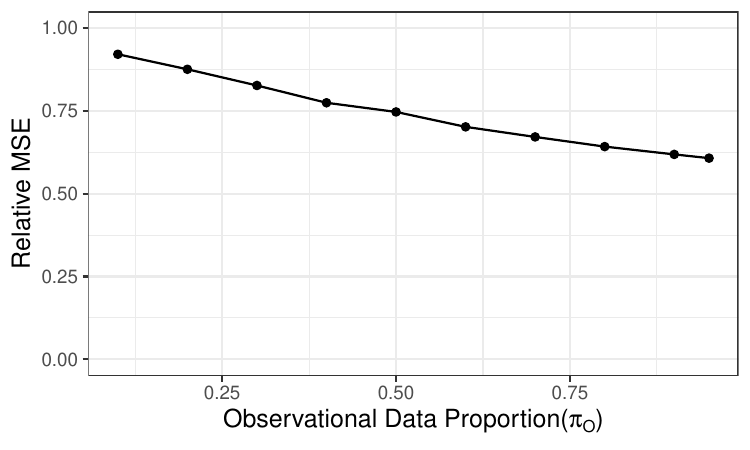}  
  \caption{Fix $R^2_{X, Z, O} = 0.6$, change $\pi_O$}
  \label{fig:relative_mse_pi}
\end{subfigure}
\begin{subfigure}{.48\textwidth}
  \centering
  \includegraphics[width=.95\linewidth]{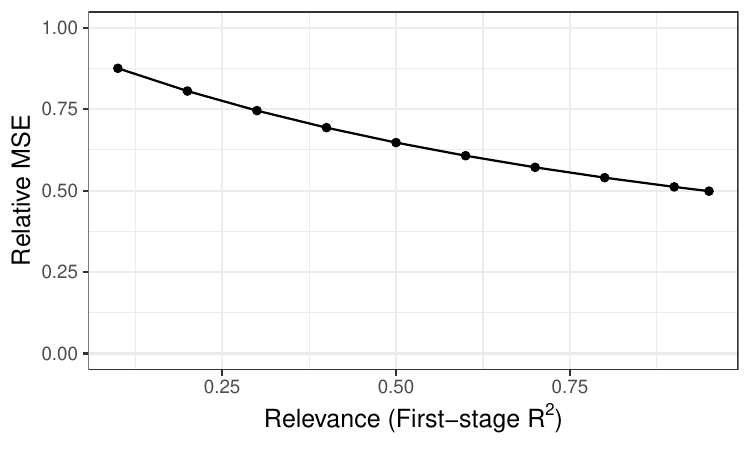}  
  \caption{Fix $\pi_O = 0.95$, change $R^2_{X, Z, O}$}
  \label{fig:relative_mse_gamma}
\end{subfigure}
\caption{Relative MSE $\frac{MSE(\widehat{\beta_1}^{Combine})}{MSE(\widehat{\beta_1}^E)}$ for different values of $\pi_O$ and $R^2_{X, Z, O}$.} 
\label{fig:relative_mse}
\end{figure}

To investigate the sensitivity of the efficiency gain,  I compare the MSE for each method under different values of $\pi_O$ and $R^2_{X, Z, O}$.  Figure \ref{fig:relative_mse_pi} shows the relative MSE for different values of $\pi_O$ while fixing the first-stage $R^2_{X, Z, O}$ to be $0.6$. 
Figure \ref{fig:relative_mse_gamma} shows the relative MSE for different values of first-stage $R^2_{X, Z, O}$ while fixing the proportion of observational data at $\pi_O = 0.95$.  This result illustrates the prediction of Lemma \ref{lemma:same_v_gain}, that the $MSE(\widehat{\beta_1}^{Combine})$ is small if the proportion of observational data is large and $Z$ is a relevant imperfect instrument.

\section{Empirical Application}\label{sec:empirical}

I apply my method to estimate the effect of website positions on clicks using the Expedia ranking dataset in \cite{ursu_power_2018}. The dataset consists of consumers’ search queries for hotels on Expedia, with each search query containing a ranked list of hotels. For each hotel in a query, I observe its rank, characteristics, and click outcome. There are $4.5$ million such observations: two thirds of the data are observational in which the hotel rankings are ordered by relevance, and one third of the data are experimental in which the hotel rankings are randomized. Consider the following model:
\begin{equation}\label{eqn:expedia_main}
Click_{i} = \alpha + f_1(Position_{i}; \theta) + \beta_2 f_2(Hotel Characteristics_{i}) + \epsilon_{i}
\end{equation}
where each observation corresponds to a queried hotel $i$ and $Position_i$ is the numerical rank of the hotel in query $i$.  To illustrate how to apply the nonlinear model,  assume the impact of position on click to be nonlinear:
$$
f_1(Position_i; \theta) =  \theta_{1} Position_i^{\theta_2}
$$
Also assume the position is affected by hotel characteristics in the observational data:
$$
Position_{i} = \begin{cases}
 f_2(HotelCharacteristics_{i}) + V_{i} & G_{i} = O\\
 Randomized & G_{i} = E
 \end{cases}.
$$
One challenge is to derive a relevant imperfect instrument $Z_i$ to predict positions that may depend on locations and other characteristics. I select locations that have more than $10,000$ observations and for each location apply a random forest procedure that uses the hotel characteristics to predict the hotel position in the observational data to estimate $f_2$:
$$
Z_{i} \equiv f_2(HotelCharacteristics_{i})
$$
The experimental and observational datasets respectively contain $N_E = 452,974$ and $N_O = 833,736$ observations in these selected locations.  Table \ref{tab:first_stage_summary} shows the observational first-stage regression in which $Z$ is a relevant covariate that explains $57.5\%$ of variation in terms of $R^2$.
\begin{table}[htbp!]
\caption{First Stage in Observational Data}\label{tab:first_stage_summary}

\begingroup 
\small 
\begin{tabular}{@{\extracolsep{5pt}}lc} 
\\[-1.8ex]\hline 
\hline \\[-1.8ex] 
 & \multicolumn{1}{c}{\textit{Dependent variable:}} \\ 
\cline{2-2} 
\\[-1.8ex] & Position \\ 
\hline \\[-1.8ex] 
 f(HotelCharacteristics) & 1.784$^{***}$ (0.002) \\ 
  Constant & $-$2.270$^{***}$ (0.021) \\ 
 \hline \\[-1.8ex] 
Observations & 833,736 \\ 
R$^{2}$ & 0.575 \\ 
Residual Std. Error & 7.027 (df = 833734) \\ 
F Statistic & 1,126,620.000$^{***}$ (df = 1; 833734) \\ 
\hline 
\hline \\[-1.8ex] 
\textit{Note:}  & \multicolumn{1}{r}{$^{*}$p$<$0.1; $^{**}$p$<$0.05; $^{***}$p$<$0.01} \\ 
\end{tabular} 
\endgroup 

\end{table}

To estimate the parameters $\Theta = \{\alpha, \theta_1, \theta_2, \beta_2\}$,  let $h_i(\Theta) =  \alpha + f_1(Position_i; \theta)+ \beta_2 f_2(Hotel Characteristics_{i})$.  The experiment-only estimate can be derived by minimizing 
\begin{align*}
\widehat{\Theta}^E & = arg\min_{\theta}\sum_{i:G_i = E}(Y_i - h_i(\Theta))^2
\end{align*}
and the combined estimate can be derived by minimizing
\begin{align*}
\widehat{\Theta}^{Combine} & = arg\min_{\theta}\sum_{i:G_i = E}(Y_i - h_i(\Theta))^2 + \lambda \left[\sum_{i:G_i = O}(Y_i - h_i(\Theta))Z_i\right]^2
\end{align*}
where I choose $\lambda = \frac{1}{Var(Z_i)}$ to mimic the optimal GMM weight in the linear model. 

To examine the effectiveness of my method,  I focus on a case when the data sample includes a large number of observational units,  with $n_O = 190,000$, and a small number of experimental units, with $n_E = 10,000$.  This sample can be simulated by randomly drawing $n_O$ units from the observational data and $n_E$ units from the experimental data.  I repeatedly draw $10,000$ such samples and perform estimation on each sample.  The resulting estimates are then compared to the approximated ground truth,  which is estimated using all experimental data and has values of $\theta_1 = 0.14$ and $\theta_2 = -0.23$.\footnote{This approximation is sufficient for evaluating MSE because the size of the original experimental dataset is more than $40$ times the size of the small experimental sample.}

Table \ref{tab:estimate_distribution} summarizes the distribution of the estimates across these $10,000$ samples.  Because of nonlinearity and limited sample size,  the distribution of experiment-only estimate for $\theta_1$ is skewed,  which may contribute to its larger variance.

\begin{table}[htbp!]
\caption{Distribution of estimated parameters across 10,000 samples}\label{tab:estimate_distribution}
\begin{tabular}{llllllllll}
\toprule
&  &  & \multicolumn{7}{c}{Percentile} \\
\cline{4-10}\\
Estimate & Mean & S.D.  & 1\% & 5\% & 25\% & 50\% & 75\% & 95\% & 99\%\\
\midrule
\multicolumn{6}{l}{\textbf{$\theta_1 = 0.14$}}\\
Experiment & 0.16 & 0.14 & 0.09 & 0.10 & 0.12 & 0.14 & 0.15 & 0.33 & 0.94\\
Combined & 0.14 & 0.03 & 0.10 & 0.11 & 0.12 & 0.13 & 0.14 & 0.16 & 0.29\\
\midrule
\multicolumn{6}{l}{$\theta_2 = -0.23$}\\
Experiment & -0.25 & 0.12 & -0.77 & -0.52 & -0.24 & -0.23 & -0.22 & -0.07 & -0.02\\
Combined & -0.24 & 0.06 & -0.59 & -0.25 & -0.24 & -0.24 & -0.23 & -0.22 & -0.08\\
\bottomrule
\end{tabular}
\end{table}

Table \ref{tab:parameter_nonlinear_comparison} compares the bias,  variance,  and MSE of 1) experiment-only estimator  2) combined estimator,  3) a nonlinear least square estimator using observational data, and 4) an IV estimator using observational data.  Both observational methods outperform the experiment-only estimator in terms of MSE, mainly due to their smaller variance. This difference highlights the importance of variance reduction, especially when sample size is limited and the model is nonlinear. In contrast,  the combined estimator has much smaller variance compared to the experiment-only estimator, resulting in a reduction of mean-squared-error by $96.4\%$ and $73.2\%$ for $\theta_1$ and $\theta_2$, respectively. The combined estimator also outperforms the two observational estimators in terms of bias and MSE for $\theta_1$; for $\theta_2$, the combined estimator has relatively less bias, while the observational methods have relatively less variance.

\begin{table}[htbp!]
  \begin{threeparttable}
    \caption{Summary statistics of estimated $(\theta_1, \theta_2)$ over $10,000$ samples}
	\label{tab:parameter_nonlinear_comparison}
\begin{tabular}{lccccc}
\toprule
Estimator & Bias$^2$ & Variance & MSE & Relative MSE & Efficiency Gain\\
\midrule
$\theta_1$\\
Experiment Only & 0.040 & 1.071 & 1.111 & 1.000 & 0.000\\
Combined & 0.000 & 0.040 & 0.040 & 0.036 & 0.964\\
NLS (Obs) & 0.422 & 0.002 & 0.424 & 0.382 & 0.618\\
IV (Obs) & 0.425 & 0.001 & 0.426 & 0.383 & 0.617\\
\midrule
$\theta_2$\\
Experiment Only & 0.005 & 0.279 & 0.284 & 1.000 & 0.000\\
Combined & 0.001 & 0.075 & 0.076 & 0.268 & 0.732\\
NLS (Obs) & 0.064 & 0.001 & 0.065 & 0.228 & 0.772\\
IV (Obs) & 0.073 & 0.000 & 0.073 & 0.257 & 0.743\\
\bottomrule
\end{tabular}
    \begin{tablenotes}
      \small
      \item Notes: $Bias^2$, Variance, and MSE are normalized by $\theta^2$ to reflect percentage error.
    \end{tablenotes}
  \end{threeparttable}
\end{table}
To assess the relative importance of bias and variance,  I summarize how well different methods estimate the marginal effect of changing position, denoted by $\beta_1$: 
$$
\beta_1\equiv E[\frac{\partial f_1(Position_i; \Theta)}{\partial Position_i}] 
$$ 
The results,  summarized in Table \ref{tab:benchmark_summary},  demonstrate that observational methods have much worse performance than the experiment-only estimate due to larger bias.  In contrast, the combined estimator has relatively smaller bias and much smaller variance,  resulting in a 45.8\% reduction in MSE.

\begin{table}[htbp!]
  \begin{threeparttable}
    \caption{Summary statistics of estimated marginal effect over $10,000$ samples}
\label{tab:benchmark_summary}

\begin{tabular}{lccccc}
\toprule
Estimator & Bias$^2$ & Variance & MSE & Relative MSE & Efficiency Gain\\
\midrule
Experiment Only & 0.000 & 0.037 & 0.037 & 1.000 & 0.000\\
Combined & 0.000 & 0.020 & 0.020 & 0.542 & 0.458\\
NLS (Obs) & 0.885 & 0.003 & 0.887 & 23.787 & -22.787\\
IV (Obs) & 0.928 & 0.002 & 0.930 & 24.921 & -23.921\\
\bottomrule
\end{tabular}

    \begin{tablenotes}
      \small
      \item Notes: $Bias^2$, Variance, and MSE are normalized by $\beta_1^2$ to reflect percentage error.
    \end{tablenotes}
  \end{threeparttable}
\end{table}

\subsection{Limitations of the Expedia Application}

This application has several limitations. As discussed in \cite{ursu_power_2018}, the dataset has some selection bias and therefore the distribution of queried hotels in the experimental dataset is different from that in the observational dataset.
 However, this selection bias only supports the robustness of the combined method: even though Assumption \ref{asmp:random_sampling} is violated because the observational and experimental units are drawn from slightly different distributions, incorporating the observational sample can still improve the precision of the experimental estimates. In addition to the distributions being different, the reduced form model in Equation \ref{eqn:expedia_main} may be misspecified in several ways. First,  the model does not impose the constraint that the outcome click must be either $0$ or $1$. Second, I assume that each observation is $i.i.d.$ but the positions and clicks may be negatively correlated within the same query.  Alleviating these concerns requires developing a rich structural model of consumer search and estimating it using a full information maximum likelihood approach. Since dealing with model misspecification and developing structural model is not the primary goal of this paper, I leave this to future extension. 

Despite these limitations, this dataset has several advantages for evaluating the effectiveness of my methods. First, the dataset contains both observational and experimental data.  Second, it contains covariates that satisfy the first-stage relevance condition. Variables, such as rating and price, are likely to affect hotel rankings. Third, the dataset is relatively large, allowing us to approximate the true causal parameter using the entire experimental data. An accurate approximation of the underlying truth allows us to evaluate parameters estimated using only a small subsample of the experimental data. Therefore,  this empirical dataset is still valuable for demonstrating the usefulness of the method.

\section{Conclusion}

This paper presents a new method that combines experimental and observational data to improve estimation efficiency.  I focus on a setting in which the experimental data are sufficient for identification but limited in size, and the observational data are large but suffer from endogeneity. The treatment of interest is randomized in the experimental data, but is endogenously determined in the observational data. I show that if the treatment is affected by or correlated with some pretreatment covariates $Z$, then these covariates can be used as imperfect instruments to derive additional moment conditions to improve estimation efficiency. 

The efficiency gain can be substantial under both linear and nonlinear models.  For linear models, I prove that the method can reduce the variance by up to $50\%$,  with the exact efficiency gain determined by the size of the observational data and the relevance of the imperfect instrument.  To illustrate the efficiency gain for nonlinear models with multiple parameters,  I apply the method to the Expedia dataset that estimates the effect of rankings on clicks.  The method reduces the MSE of some parameters by as high as $97\%$ and reduces the MSE of the overall marginal effect by approximately $46\%$.

Increasing the estimation efficiency is beneficial for firms for several reasons.  From a data analysis perspective, increasing the estimation efficiency means that firms can leverage existing datasets to obtain a more precise measurement of an effect, which in turn increases the likelihood of making the correct managerial decision.  From an experimental design perspective,  higher efficiency means that firms can reduce the required experimental sample size to detect a given effect.  For example,  a $50\%$ reduction in variance means that firms can reduce the required experimental sample size by $50\%$.  This reduction in experimental size is valuable because randomization in experiments may negatively impact user experience by offering suboptimal services or products.  Additionally,  firms often have a limited number of customers at any given moment,  so reducing the required experiment size means that firms can reduce the required time for completing an experiment,  which speeds up the learning and decision-making process. 

In addition to the application on Expedia ranking,  the method of combining datasets can be extended to other settings in which firms observe direct or indirect inputs to algorithms that affect the treatment.  For example,  past consumption may affect targeted advertising and coupon,  customer credit score may affect credit loan interest rate,\footnote{Appendix B.5.1 
discusses this application using semi-simulated data from \cite{bertrand_whats_2010}.} age may affect insurance premium,  and demand forecast may affect prices.  In these scenarios,  firms can leverage the observational data to improve the accuracy of the experimental estimate and speed up learning.

My method is not without limitations. It is most useful when the unobserved error enters the outcome model additively, and when its distributions are similar in the experimental and observational samples. These two conditions allow additional moment restrictions to be derived from the observational data. When these conditions are satisfied, my method strictly decreases the variance without introducing significant bias. When these assumptions are moderately violated, the combined estimator can be thought of as a more efficient estimator that may be biased due to misspecified moment conditions.\footnote{Appendix B.4 
 discusses how to account for model misspecifications.} Although my method may introduce additional bias in this scenario, my method can still improve the overall MSE by reducing the variance.

\newpage

\appendix
\section{Proof}

\subsection{Proof of Lemma \ref{lemma:uncorrelation} and \ref{lemma:inverse_weighting}}\label{proof:uncorrelation}

\textbf{Lemma \ref{lemma:uncorrelation}}: $\widehat{\beta_1}^O$ and $\widehat{\beta_1}^E$ are uncorrelated.\\
\textbf{Proof}:
Since $(\widehat{\beta_1}^O,\widehat{\gamma}^O)$ are estimated using observational data while $(\widehat{\beta_1}^E, \widehat{b_2}^E)$ are estimated using experimental data, these pairs of estimates are independent:
\begin{equation}\label{proof:uncorrelation:eqn1}
(\widehat{\beta_1}^O, \widehat{\gamma}^O) \perp (\widehat{\beta_1}^E, \widehat{b_2}^E) 
\end{equation}
Because $X_i$ and $Z_i$ in the experimental data are independent and are used to estimate $(\widehat{\beta_1}^E, \widehat{b_2}^E)$:
\begin{equation}\label{proof:uncorrelation:eqn2}
Cov(\widehat{\beta_1}^E, \widehat{b_2}^E) = 0.
\end{equation}
Plugging Equations \ref{proof:uncorrelation:eqn1} and \ref{proof:uncorrelation:eqn2} into the covariance:
$$
\begin{aligned}
Cov(\widehat{\beta_1}^O, \widehat{\beta_1}^E) & = Cov(\widehat{b_1}^{IV} - ({\widehat{b_2}^E}/{\widehat{\gamma}^O}),  \widehat{\beta_1}^E)\\
& = Cov(\widehat{b_1}^{IV},  \widehat{\beta_1}^E) -  Cov(\frac{ \widehat{b_2}^E}{\widehat{\gamma}^O}, \widehat{\beta_1}^E) \\
& = Cov(\widehat{b_1}^{IV},  \widehat{\beta_1}^E) - E[\frac{1}{\widehat{\gamma}^O}] Cov( \widehat{b_2}^E, \widehat{\beta_1}^E) \\
& = 0
\end{aligned}
$$

\noindent \textbf{Lemma \ref{lemma:inverse_weighting}}: Under Lemma \ref{lemma:uncorrelation} the weighting is optimal if each estimate is weighted in inverse proportion to its variance:
$$
\frac{w_O^*}{w_E^*} = \frac{Var(\widehat{\beta_1}^E)}{Var(\widehat{\beta_1}^O)}
$$\\
\textbf{Proof:}
\begin{equation}
\begin{aligned}
Var(w_O \widehat{\beta_1}^O + w_E \widehat{\beta_1}^E) & = w_O^2 Var(\widehat{\beta_1}^O) + w_E^2 Var(\widehat{\beta_1}^E) + 2 w_O w_E Cov(\widehat{\beta_1}^O, \widehat{\beta_1}^E)\\
& = w_O^2 Var(\widehat{\beta_1}^O) + w_E^2 Var(\widehat{\beta_1}^E) 
\end{aligned}
\end{equation}
Taking the first order condition subject to the constraint that $w_O + w_E = 1$:

$$
w_O = \frac{Var(\widehat{\beta_1}^E)}{Var(\widehat{\beta_1}^E) + Var(\widehat{\beta_1}^O)}
$$

$$
w_E = \frac{Var(\widehat{\beta_1}^O)}{Var(\widehat{\beta_1}^E) + Var(\widehat{\beta_1}^O)}
$$

\subsection{Proof of Lemma \ref{lemma:epsilon}}\label{proof:epsilon}

Recall that $b_2$ is defined in equation \ref{eqn:b2} as 
$$
b_2 \equiv \frac{Cov(Y_i - \beta_1 X_i, Z_i|G_i = O)}{Var(Z_i|G_i = O)} = \beta_2 + \frac{Cov(U_i, Z_i|G_i = O)}{Var(Z_i|G_i = O)}
$$
Then 
\begin{align*}
\epsilon_i & = Y_i - \beta_1 X_i - b_2 Z_i \\
& = Y_i - \beta_1 X_i -  (\beta_2 + \frac{Cov(U_i, Z_i|G_i = O)}{Var(Z_i|G_i = O)}) Z_i \\
& = (Y_i - \beta_1 X_i -  \beta_2 Z_i) - \frac{Cov(U_i, Z_i|G_i = O)}{Var(Z_i|G_i = O)} Z_i \\
& = U_i - \frac{Cov(U_i, Z_i|G_i = O)}{Var(Z_i|G_i = O)} Z_i 
\end{align*}
In the observational data:
\begin{align*}
Cov(\epsilon_i, Z_i|G_i = O) & = Cov\left(U_i - \frac{Cov(U_i, Z_i|G_i = O)}{Var(Z_i|G_i = O)}Z_i, Z_i|G_i = O\right) \\
& = Cov(U_i, Z_i|G_i = O) - \frac{Cov(U_i, Z_i|G_i = O)}{Var(Z_i|G_i = O)}Cov(Z_i, Z_i|G_i = O)\\
& = Cov(U_i, Z_i|G_i = O) - Cov(U_i, Z_i|G_i = O) \\
& = 0
\end{align*}
Given Assumption \ref{asmp:random_sampling}
$$
\epsilon_i = U_i - \frac{Cov(U_i, Z_i|G_i = O)}{Var(Z_i|G_i = O)}Z_i = U_i - \frac{Cov(U_i, Z_i|G_i = E)}{Var(Z_i|G_i = E)}Z_i
$$
so the same relationship holds for the experimental data: $Cov(\epsilon_i, Z_i|G_i = E) = 0$.

\subsection{GMM Estimator}\label{appendix:gmm_analytical}

This section derives an GMM estimator for the general multivariate case using matrix notation.  It is convenient to re-write the moment conditions and the minimization problem into matrix notation. 
Let 
$$
A = \begin{bmatrix}
\mathbf{X} & \mathbf{Z}\\
\end{bmatrix}
$$
denote the matrix of independent variables.  For simplicity, let  
$$
B = 
\begin{bmatrix}
\mathbf{X}_E & \mathbf{Z}_E  & 0 \\
0 & 0  & \mathbf{Z}_O
\end{bmatrix}
$$
be the matrix that corresponds to the two conditions in the experimental data and one condition in the observational data, and 
$$
\theta = 
\begin{bmatrix}
\beta_1 \\
b_2\\
\end{bmatrix}
$$
be the vector parameter of interest. 
Then 
$$
\frac{1}{N} \sum_{i = 1}^N g_i(\widehat{\beta_1}, \widehat{b_2}) = \frac{1}{N} (B' (Y - A\theta))
$$
and the GMM with a weighting matrix $W$ minimizes:
$$
(Y - A\theta)'B W B' (Y - A\theta)
$$
After taking the first order conditions with respect to $\theta$, one can get:
$$\widehat{\theta} = (A'B W B' A)^{-1}(A'B W B'Y).$$

\subsection{Proof of Theorem \ref{thm:basic}}\label{proof:theorem_gmm}

Recall that the moment function is 
$$
g_i(\beta_1, b_2) = 
\begin{bmatrix}
(Y_i - X_i \beta_1 -Z_i b_2  ) X_i I(G_i = E) \\
(Y_i - X_i \beta_1  - Z_i b_2  ) Z_i I(G_i = E)\\
 (Y_i - X_i \beta_1  - Z_i b_2  ) Z_i I(G_i = O)
\end{bmatrix}
$$
The covariance matrix of the moment function $\Omega$ is:
$$
\begin{aligned}
\Omega & = E\left[g_i(\beta_1, b_2) g_i(\beta_1, b_2) ^T\right] \\
& = \begin{bmatrix}
\pi_{E} E[ \epsilon_i^2 X_i'X_i |G_i = E] & \pi_{E} E[ \epsilon_i^2 X_i'Z_i |G_i = E]  & 0 \\
\pi_{E} E[ \epsilon_i^2 Z_i'X_i |G_i = E]  & \pi_{E} E[ \epsilon_i^2 Z_i'Z_i |G_i = E] & 0 \\
0 & 0 & \pi_{O} E[ \epsilon_i^2  Z_i'Z_i |G_i = O] 
\end{bmatrix}	
\end{aligned}
$$
which can be simplified under Assumption \ref{asmp:no_homo_auto}:
$$
\begin{aligned}
\Omega & = \frac{1}{N}\begin{bmatrix}
X_E'X_E & X_E'Z_E& 0 \\
Z_E'X_E & Z_E'Z_E  & 0 \\
0 & 0 & Z_O'Z_O\\
\end{bmatrix} \sigma^2 
\end{aligned}
$$
Let $\Gamma$ be matrix of the derivatives of the moment functions with respect to $(\beta_1, b_2)$:
$$
\begin{aligned}
\Gamma & = \frac{1}{N}\begin{bmatrix}
\frac{\sum_i g_{i1}}{\partial \beta_1} & \frac{\sum_i g_{i1}}{\partial b_2}\\
\frac{\sum_i g_{i2}}{\partial \beta_1} & \frac{\sum_i g_{i2}}{\partial b_2}\\
\frac{\sum_i g_{i3}}{\partial \beta_1} & \frac{\sum_i g_{i3}}{\partial b_2}
\end{bmatrix}  = \frac{1}{N} \begin{bmatrix}
X_E'X_E & X_E'Z_E\\
Z_E'X_E & Z_E'Z_E \\
Z_O'X_O & Z_O'Z_O 
\end{bmatrix}
\end{aligned}
$$
Note that $\frac{X_E'Z_E}{N_E} \to 0$ because of randomization.  Under an efficiency weight matrix $W = \Omega^{-1}$,  the asymptotic variance can be calculated based on the sandwich formula:
$$
\begin{aligned}
V(\widehat{\beta_1}^{GMM}, \widehat{b}_2) & = \frac{1}{N}(\Gamma' \Omega^{-1} \Gamma)^{-1} \\
& = \sigma^2 \begin{bmatrix}
X_E'X_E + X_O'Z_O (Z_O'Z_O)^{-1} Z_O'X_O  & X_O'Z_O\\
Z_O'X_O & Z_E'Z_E + Z_O'Z_O\\
\end{bmatrix}^{-1}\\
\end{aligned}
$$
With block inversion, the variance of $\widehat{\beta_1}^{GMM}$ can be written as
$$
\begin{aligned}
V(\widehat{\beta_1}^{GMM}) & = \sigma^2 [X_E'X_E + X_O'Z_O (Z_O'Z_O)^{-1} Z_O'X_O  - X_O'Z_O(Z_E'Z_E + Z_O'Z_O)^{-1}Z'_O X_O ]^{-1}\\
& = \sigma^2 [X_E'X_E + X_O'Z_O \left( (Z_O'Z_O)^{-1} -(Z_E'Z_E + Z_O'Z_O)^{-1}\right)  Z_O'X_O ]^{-1}\\
\end{aligned}
$$
When both $X_i$ and $Z_i$ are univariate random variables with $0$ means, the variance can be simplified to 
$$
\begin{aligned}
V(\widehat{\beta_1}^{GMM}) &  = \sigma^2 \left[n_E Var(X_i|G_i = E) + \frac{n_O^2 Cov(X_i, Z_i|G_i = O)^2 n_E Var(Z_i|G_i = E)}{(n_E + n_O)n_O Var(Z_i) Var(Z_i|G_i = O)}\right]^{-1}\\
&  = \sigma^2 \left[n_E Var(X_i|G_i = E) + \frac{n_O \gamma^2 Var(Z_i|G_i = O)^2 n_E Var(Z_i|G_i = E)}{(n_E + n_O)Var(Z_i) Var(Z_i|G_i = O)}\right]^{-1}\\
&  = \frac{\sigma^2}{n_E} \left[Var(X_i|G_i = E) + \pi_O \frac{Var(\gamma Z_i|G_i = O) Var(Z_i|G_i = E)}{Var(Z_i)}\right]^{-1}\\
\end{aligned}
$$
When Assumption \ref{asmp:random_sampling} is satisfied such that $Var(Z_i|G_i = E) = Var(Z_i|G_i = O) = Var(Z_i)$. The variance can be further simplified to 
$$
\begin{aligned}
V(\widehat{\beta_1}^{GMM}) 
&  = \frac{\sigma^2}{n_E} \left[Var(X_i|G_i = E) + \pi_O Var(\gamma Z_i|G_i = O)\right]^{-1}\\
\end{aligned}
$$
This completes the proof of Theorem \ref{thm:basic}.

\newpage

\section{Extensions}

\subsection{Extension to Conditional Similarity}\label{sec:extension_conditional_independence} 

Assumption \ref{asmp:random_sampling} assumes that the distributions of characteristics are identical between experimental and observational data.  This assumption is plausible if the experimental unit is randomly selected.  This assumption can be relaxed to assume that the distributions are identical after adjusting for observed covariate $Z_i$:

\begin{assumption}[Conditional Similarity in Unobservables]\label{asmp:similarity}
$$
G_i \perp U_i|Z_i
$$
\end{assumption}
This conditional independence condition is common in the related literature (e.g., \cite{athey_combining_2020}, \cite{kallus_removing_2018}) to ensure effects learned in one dataset can be transferred to another dataset. 

Compared to Assumption \ref{asmp:random_sampling}, this relaxed assumption allows for the possibility of stratified sampling in which the probability of selecting into the experimental dataset depends on observed covariates.  For example, firms may choose to reduce the experimental traffic during certain business hours.  If firms believe that experimenting with new customers is relatively riskier,  firms may also consider reducing the amount of experimental traffic for relatively new customers.  In this case,  Assumption \ref{asmp:similarity} holds if $Z$ includes the variables firms use for stratification.

My method can be extended to account for this relaxed assumption by re-weighting or re-sampling the observational data such that the distribution of $Z$ for these two datasets match. 
 
\subsection{Nonlinear Model Efficiency Gain: A GMM Perspective}\label{sec:nonlinear_efficiency}

Section \ref{sec:extension_nonlinear} takes a constraint optimization perspective to discuss why the efficiency gain may be increased in nonlinear models.  This section takes a GMM perspective to discuss why efficiency gain can be improved under the nonlinear model.  For illustration,  consider the case when $Z$ is a continuous covariate observed by researcher, $X$ is a continuous variable, and the function $h$ is smooth in the continuous $X$ and $Z$,  such that the partial derivatives of the function $h$ are well defined.\footnote{Additional parametric assumptions may be needed for binary $Z_i$ and $X_i$.}

Consider the neighborhood of any $(X_0, Z_0)$. Let $b_2 \equiv \frac{\partial h(X_0, Z_0; \theta)}{\partial Z}$, and $\beta_1 \equiv \frac{\partial h(X_0, Z_0; \theta)}{\partial X}$. Then around the neighborhood of $(X_0, Z_0)$:

$$
\begin{aligned}
h(X_i, Z_i; \theta) & \approx h(X_0, Z_0; \theta) + \frac{\partial h(X_0, Z_0; \theta)}{\partial X} (X_i - X_0) + \frac{\partial h(X_0, Z_0; \theta)}{\partial Z} (Z_i - Z_0) \\
& =  h(X_0, Z_0; \theta) + \beta_1 (X_i - X_0) + b_2 (Z_i - Z_0) \\
& = h(X_0, Z_0; \theta) -  \beta_1 X_0 - b_2  Z_0 +  \beta_1 X_i + b_2 Z_i
\end{aligned}
$$
Let $\alpha_0 = h(X_0, Z_0; \theta) - \beta_1 \cdot X_0 - b_2 \cdot Z_0$ be the local intercept such that 
$$
h(X_i, Z_i; \theta) \approx \alpha_0 + \beta_1 X_i + b_2 Z_i
$$
Then we can derive local moment conditions from experimental data:
$$
\begin{aligned}
E[(Y_i - \alpha_0  - \beta_1X_i - b_2 Z_i) 1|G_i = E] \approx 0\\
E[(Y_i - \alpha_0  - \beta_1X_i - b_2 Z_i) X_i|G_i = E] \approx 0\\
E[(Y_i - \alpha_0  - \beta_1X_i - b_2 Z_i) Z_i|G_i = E] \approx 0\\
\end{aligned}
$$
as well as additional moment conditions from the observational data
$$
\begin{aligned}
E[(Y_i - \alpha_0  - \beta_1X_i - b_2 Z_i)1 |G_i = O] \approx 0\\
E[(Y_i - \alpha_0  - \beta_1 X_i - b_2 Z_i) Z_i|G_i = O] \approx 0\\
\end{aligned}
$$
Because these local moment conditions are similar to the moment conditions in the linear case, the accuracy of $\beta_1$ can also be improved based on Theorem \ref{thm:basic}, where the determinants of efficiency gain is the local variance and first-stage relevance of $Z_i$, as well as the proportion of observational data.  If the accuracy of $\beta_1$ is improved,  then by definition the accuracy for $\frac{\partial h(X, Z; \theta)}{\partial X}$ will also be improved,  which means that we can better estimate the marginal impact of X.  

\subsection{Extension to Binary Probit}\label{sec:probit}

In this section, I illustrate how the observational data can be incorporated into the case of binary probit with endogenous and continuous regressors following a textbook example in Wooldridge (2010, Section
15.7.2), where:
\begin{equation}\label{eqn:y_nonlinear}
Y_i = \begin{cases}
1 & Y_i^*> 0\\
0 & \text{otherwise}
\end{cases}
\end{equation}
where $Y_i^* = \beta_1 X_i + \beta_2 Z_i + U_i$.  I hold the first-stage equation to be the same as in Equation \ref{eqn:x}:
$$
X_i = \begin{cases} 
\gamma Z_i + V_i & \text{ if }G_i = O\\
randomized & \text{ if } G_i = E\\
\end{cases}. 
$$
The standard probit assumes that $(U_i, V_i)$ follows a standard normal distribution that is independent of the instrumental variable $Z_i$. To make the case more general, I allow $Z_i$ to be correlated with $U_i$:\footnote{I maintain the assumption that $Cov(Z_i, V_i|G_i = O) = 0$ since $\gamma$ is a linear projection parameter,  Since I assume $X_i$ is determined by ($Z_i$, $V_i$) and I do not consider simultaneous equation, I do not allow the distribution to depend on $X_i$}
$$
(U_i, V_i|Z_i) \sim N 
\left(\begin{bmatrix}
\rho_{zu} \times Z_i  \\
0 \\
\end{bmatrix}, 
\begin{bmatrix}
1 & \rho_{uv} \sigma_v\\
\rho_{uv} \sigma_v & \sigma_v^2
\end{bmatrix}
\right)
$$
$Z_i$ is therefore not an instrumental variable because it either directly affects $Y_i^*$, or correlated with unobservables $U_i$. 

As in my linear case, not all parameters can be separately identified in this setup and I need to aggregate some parameters for estimation convenience. I define $b_2$ as
$$
b_2 \equiv \beta_2 + \rho_{zu} 
$$ 
which captures both the direct effect of $Z$ on $Y^*$ and an indirect correlation with the unobservables $U$. Let $\Theta$ denote the set of unknown parameters:
$$
\Theta = \{\beta_1, b_2, \rho_{uv}, \gamma, \sigma_v\}
$$ 
A maximum likelihood estimator that only uses the experimental data chooses $(\beta_1, b_2)$ to maximizes:
\begin{align*}
\ln \mathcal{L}_E(\Theta)  & = \ln \prod_{i:G_i = E} P(Y_i|X_i, Z_i, \beta_1, b_2)
\end{align*}
The log-likelihood of the observational data can be written as
\begin{align*}
\ln \mathcal{L}_O(\Theta)  & = \ln \prod_{i:G_i = O} P(Y_i, X_i|Z_i, \Theta)\\
& = \ln \prod_{i:G_i = O} P(Y_i|Z_i, X_i, \Theta)P(X_i|Z_i, \Theta)
\end{align*}
The data can be combined by optimizing the sum of these two log-likelihoods:
$$
\max_{\Theta}  \ln \mathcal{L}_E(\Theta) + \ln \mathcal{L}_O(\Theta)
$$
To understand why incorporating the observational data can help improve efficiency in this probit case, consider when the size of the observational data is infinite. $(\gamma, \sigma_v)$ are identified using the first-stage equation, where
$$
P(X_i|Z_i, G_i = O,  \Theta) = \phi(\frac{X_i - \gamma Z_i}{\sigma_v})
$$
Following Wooldridge, under normality, the second stage can be written as:
\begin{equation*}
\begin{split}
P(Y_i = 1|X_i, Z_i, G_i = O,  \Theta) & = \Phi\left[\frac{\beta_1 X_i + b_2 Z_i + \frac{\rho_{uv}}{\sigma_v}(X_i - \gamma Z_i)}{\sqrt{1 - \rho_{uv}^2}}\right]
\\
& = \Phi\left[\frac{1}{\sqrt{1 - \rho_{uv}^2}}(\beta_1 + \frac{\rho_{uv}}{\sigma_v})X_i +\frac{1}{\sqrt{1 - \rho_{uv}^2}}( b_2 - \frac{\rho_{uv}}{\sigma_v} \gamma)Z_i \right]\\
\end{split}
\end{equation*}
A probit of $Y_i$ on $(X_i, Z_i)$ on observational data can then consistently identify two constraints:
\begin{align*}
C_1 = \frac{1}{\sqrt{1 - \rho_{uv}^2}}(\beta_1 + \frac{\rho_{uv}}{\sigma_v})\\
C_2 = \frac{1}{\sqrt{1 - \rho_{uv}^2}}( b_2 - \frac{\rho_{uv}}{\sigma_v} \gamma)
\end{align*}
Since $(\gamma, \sigma_v)$ are already identified in the first-stage, $(b_2, \beta_1, \rho_{uv})$ are the three unknowns left. If $b_2 = 0$, the problem is reduced to an endogenous binary probit with $Z_i$ being an instrumental variable; then $(\beta_1, \rho_{uv})$ are just identified under $C_1$ and $C_2$. However, because $b_2$ is unknown, the observational data alone are not sufficient for identification. 

Although the observational data are not sufficient for identification, it can help improve efficiency of the experimental data by imposing additional constraints onto the maximum likelihood problem:
\begin{align*}
& \max_{\beta_1, b_2, \rho_{uv}} \ln \mathcal{L}_E(\beta_1, b_2)\\
\text{subject to } & C_1 = \frac{1}{\sqrt{1 - \rho_{uv}^2}}(\beta_1 + \frac{\rho_{uv}}{\sigma_v})\\
& C_2 = \frac{1}{\sqrt{1 - \rho_{uv}^2}}( b_2 - \frac{\rho_{uv}}{\sigma_v} \gamma)
\end{align*}
Intuitively, when the two constraints $C_1$ and $C_2$ are precisely measured, any deviations from these constraints must be attributed to the inaccuracy of $(\beta_1, b_2)$. Incorporating the observational data can therefore help improving the efficiency.

\subsection{Extension to Misspecified Observational Model}\label{sec:misspecification}

My method can be generalized to settings when the experimental data are still sufficient for identification, but the observational model is misspecified. Recall that Equation \ref{eqn:linear} assumes a model with additive separability that rules out any interactions. Equation \ref{eqn:parametric_nonlinear} assumes a more flexible nonlinear model with additive unobservables. Assumption \ref{asmp:similarity} assumes that distributions of unobservables are similar between observational and experimental units after conditioning on covariates. To alleviate the concerns that these assumptions may be violated in practice, I discuss strategies that help researchers determine: 1) whether they should incorporate the observational data and 2) how to tune the hyperparameter when the model may be misspecified. The method is still valuable if the goal is to minimize mean-squared error (MSE), which does not only include the bias but also the variance. 

One strategy is to use cross validation, a popular procedure for tuning hyperparameters in the regularized regression. The key is to cross validate on out-of-sample experimental data. Intuitively, a better causal estimator should more accurately predict $Y$ given a randomized $X$. Researchers can therefore test the usefulness of a combined estimator $\widehat{\beta_1}^{combine}$ relative to the experiment-only estimator $\widehat{\beta_1}^{E}$ using out-of-sample experimental data. 

Another strategy is to average multiple GMM estimators given the recent development in the literature. The key is to recognize that two estimators can be generated when both observational and experimental estimates are available:\footnote{I thank the anonymous associate editor for this valuable insight.} 
1) a valid experiment-only estimate $\hat{\beta_1}^E$ that does not require the additional assumptions, and 2) a more efficient estimator $\widehat{\beta_1}^{combine}$ that is potentially biased due to model misspecification. This setting fits into the classical Hausman specification test (\cite{hausman_specification_1978}), where researchers can decide to use the more efficient combined estimator if it is not significantly different from the less efficient one that only uses the experimental data. This setting also fits into the more recent framework developed by \cite{cheng_uniform_2019} that averages two GMM estimators, where one is always consistent, and the other one could be inconsistent due to misspecifications of moment conditions. The framework guarantees that given certain weights, the mean-squared-error can be improved. Following their framework, one can combine the two estimators:

$$
\widehat{\beta}_1^{combine, robust} = w \widehat{\beta_1}^{combine} + (1 - w) \hat{\beta_1}^E
$$
where the optimal weight $w$ is determined by the relative variance and bias differences.
\begin{align*}
w^*  & = \frac{\mathbb{V}(\hat{\beta}^{E}_1)  - \mathbb{V}(\widehat{\beta_1}^{combine})}{(E[\hat{\beta}^{E}_1] - E[\widehat{\beta_1}^{combine}])^2 + \mathbb{V}(\hat{\beta}^{E}_1)  - \mathbb{V}(\widehat{\beta_1}^{combine})}\\
& = \frac{ImprovedVariance}{Bias^2 + ImprovedVariance}
\end{align*}

Therefore, when the combined estimator is consistent, the optimal weight is approximately $1$, suggesting that one can safely use the combined estimate. When the combined estimator has large bias relative to the variance improvement, $w$ approaches $0$, suggesting that one should avoid using the combined estimate. When the bias is small, researchers can still use the combined method to improve the overall MSE. 

\subsection{Extension to Other Empirical Settings}

This section discusses additional examples under which my method could be useful.  A common reason for firms to analyze data in the past is to improve decisions and profit in the future. Firms are only interested in conducting such analyses if past customers are not too different from future customers, otherwise insights gained from the past are not useful for the future. For this type of firms, if the observational data include all customers that arrive in the past, and the experimental data include all customers that arrive in the near future, it is plausible to assume that these two populations are similar. To improve future decisions, firms need to determine the size of the experiment as well as what to do after they obtain the experimental results.

My method can improve the profit by lowering the size of the experiment needed to achieve certain statistical accuracy. According to \cite{feit_test_2019}, the profit depends on the size of the experimental sample because larger experiments imply that more customers receive suboptimal treatments under randomization, leading to lower profit. Because my method can reduce the variance by up to $50\%$, only half of the experimental sample is required to attain the same statistical precision, thus improving the profit.

\subsubsection{Credit Loan}\label{appendix:credit_loan}
\begin{figure}[H]
\centering
\includegraphics[scale=0.5]{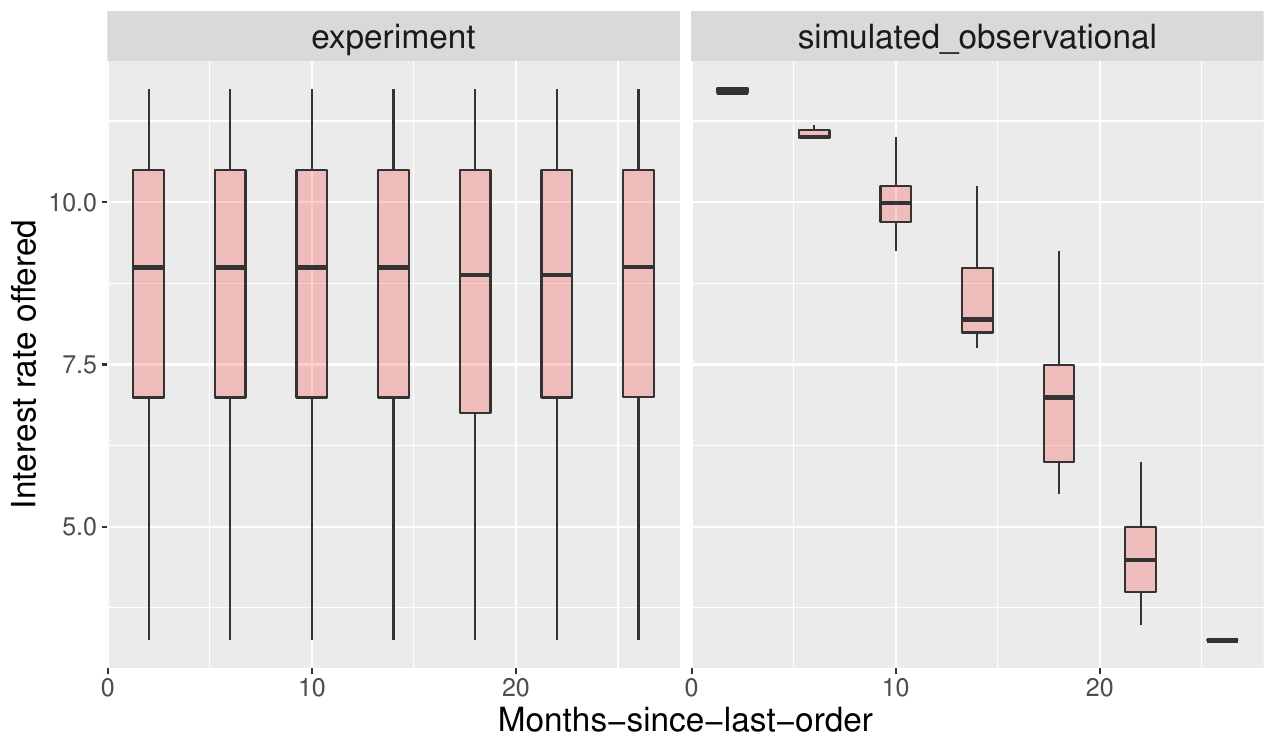}
\caption{Distribution of interest rate vs months since last order}\label{fig:advertising_boxplot}
\end{figure}

In general, it is difficult to find a dataset where both experimental and observational data are available in other contexts. To best demonstrate the efficiency gain of my method in another context, I simulated a semi-realistic setting based on a direct mail field experiment (\cite{bertrand_whats_2010}) implemented by a consumer lender that randomizes creative content and interest rates. One parameter of interest in the experiment is how the offered interest rate affects the customer loan amount, an important managerial input that firms can use to optimize future offering of interest rates. \cite{bertrand_whats_2010} assume a linear causal model to answer this question:\footnote{The OLS specification in Table (3) Column 4 of the paper.}

\begin{equation}\label{eqn:advertising_ground_truth}
\text{LoanAmount}_i = \beta_1 \text{InterestRate}_i + \beta_2 \text{ConsumerCharacteristics}_i + \epsilon_i
\end{equation}
where consumer characteristics include variables such as credit risk as well as months-since-last-loan, $\beta_2$ is the vector of corresponding coefficient, and $\beta_1$ is the parameter of interest. The experiment is sufficient for identifying the impacts of interest rate on loan because the interest rate is randomly drawn from a distribution after stratifying on credit risk.\footnote{For illustrative purposes, I focus on the customers in the high-risk stratum, whose interest rate is randomly drawn from the same distribution because the experiment is stratified. My method can be easily extended to other strata by combining experimental and observational data within those strata.} However, in practice the interest rate is directly affected by many consumer characteristics. For example, the firm may give a lower interest rate to churned customers who have not made a loan for a few months. If researchers also have access to the large amount of business-as-usual direct mail data before the experiment, then my method can incorporate this observational data to improve the efficiency of the experiment estimate. Because \cite{bertrand_whats_2010} does not include such observational data, additional assumptions are needed to to simulate such a semi-realistic setting. I first follow Equation \ref{eqn:advertising_ground_truth} to estimate the model using all the experimental sample and take the estimated parameters and unobservables as ground truth. I then assume that the interest rate is determined by a weighted average between the months-since-last-loan and unobservables, where months-since-last-loan receive $95\%$ of the weight.\footnote{In practice, many other features are used for determining the interest rate offer, such as a continuous credit score. All these variables can be incorporated as well}
 Figure \ref{fig:advertising_boxplot} shows that the interest rate is random in the experimental data, but is negatively correlated with the months-since-last-order in the simulated observational data. 

Similar to the Expedia application, I examine the effectiveness of my method when units are randomly assigned into the experimental and observational groups. I focus on a case when the observational group is large with $20,000$ units and the experimental group is small with $1,000$ units. I repeatedly draw $M = 10,000$ such samples and perform estimation on each sample. Table \ref{tab:advertising_mse_summary} reports the bias, variance, and MSE of different estimators across 10,000 replications. To characterize the efficiency gain, I compare the MSE of these methods with a benchmark that uses only a random sample of experimental data for estimation. The OLS using observational data has large bias due to endogeneity. The incorrect IV approach also has large bias, because the variable $Z_i$ violates the exclusion restriction and is not a valid IV. In comparison, the combined GMM method has a efficiency gain of $45.7\%$.

\begin{table}[htbp!]
\small
  \begin{threeparttable}
    \caption{Summary for different estimators when units are randomly assigned into experimental and observational groups}
	\label{tab:advertising_mse_summary}
	
\begin{tabular}{lccccc}
\toprule
Estimator & Bias$^2$ & Variance & MSE & Relative MSE & Efficiency Gain\\
\midrule
Experiment Only & 0.007 & 16.796 & 16.803 & 1.000 & 0.000\\
Combined GMM & 0.002 & 9.127 & 9.129 & 0.543 & 0.457\\
OLS (Obs) & 12107.454 & 56.715 & 12164.169 & 723.928 & -722.928\\
IV (Obs) & 105.028 & 0.524 & 105.552 & 6.282 & -5.282\\
\bottomrule
\end{tabular}

  \end{threeparttable}
\end{table}

For the two unbiased estimators, Table \ref{tab:advertising_significance_summary} reports how often the increased efficiency changes the statistical significance at the 95\% level. The combined GMM approach is more likely to detect that the effect of interest rate on loan is negative, and also more likely to detect that this effect is statistically significant. 

\begin{table}[!htbp] \centering 
  \caption{Summary of statistical significance  ($p < 0.05$) over $10,000$ samples} 
  \label{tab:advertising_significance_summary} 

\begin{tabular}{lll}
\toprule
Method & Negative (Correct Sign) & Significantly Negative\\
\midrule
Experiment Only & 80.38\% & 13.99\%\\
Combined GMM & 87.61\% & 21.18\%\\
\bottomrule
\end{tabular}

\end{table}

\subsubsection{Incentives}

Rideshare platforms such as Uber and Lyft are often interested in how workers respond to wages and incentives. Causally measuring the elasticity of labor supply is crucial for satisfying consumer demand. The experimental approach is to randomly change the wages. However, because randomizing wages is controversial and may lead to public backlash, firms are sometimes only willing to conduct a small-scale experiment to minimize the risk. For example, \cite{chen_value_2019} mentioned that Uber has conducted some randomized wage experiments on a limited basis in several cities. One such experiment is conducted in Orange County, California that covers approximately $3,000$ Uber drivers during April 2016. In the experiment, a random set of drivers receive an email indicating that the driver would receive a $10$ percent increase in wages for 3 weeks. \cite{chen_value_2019} use this small-scale experiment to show that the results obtained using a larger observational dataset that covers $200,000$ drivers are valid.\footnote{An implicit assumption behind this exercise is that observational and experimental population are similar, satisfying my Assumption \ref{asmp:random_sampling}}

In this setting, the experimental dataset is small, containing one market for three experimental weeks, and the observational dataset is large, containing all other non-experimental markets and weeks. In the experimental data, the wage is set randomly; in the observational data, the wage is affected by many factors observed by firms, such as local consumer demand and worker characteristics. Because workers' willingness to work may be correlated with these factors, observational methods that incorrectly use these factors as IV may generate biased results. Because my method can account for this bias, researchers can derive an additional bias-corrected estimate that is uncorrelated with the experiment-only estimate to improve the efficiency. 

\newpage

\bibliographystyle{apalike}
\bibliography{ExperimentValue.bib}

\end{document}